\documentclass[fleqn,usenatbib]{mnras}
\usepackage{newtxtext,newtxmath}
\usepackage[T1]{fontenc}
\usepackage{ae,aecompl}
\usepackage{xcolor}

\usepackage{graphicx}	
\usepackage{amsmath}	
\usepackage{subcaption}
\usepackage{bm}
\usepackage[utf8]{inputenc}

\newcommand{\alfB}{\alpha_{_{\rm B}}(T)}
\newcommand{\alfBO}{\alpha_{_{\rm B}}}
\newcommand{\BO}{B_{_{\rm O}}}
\newcommand{\cII}{c_{_{\rm II}}}
\newcommand{\cLAY}{c_{_{\rm LAYER}}}
\newcommand{\cSUR}{c_{_{\rm SUR}}}
\newcommand{\cSURsqd}{c_{_{\rm SUR}}^2}
\newcommand{\delSUR}{\delta_{_{\rm SUR}}(\tau)}
\newcommand{\dMIIdt}{dM_{_{\rm II}}/dt}
\newcommand{\dMSURdt}{\frac{dM_{_{\rm SUR}}}{dt}}
\newcommand{\dWIFdt}{dW_{_{\rm IF}}/dt}
\newcommand{\dWIFdtX}{\frac{dW_{_{\rm IF}}}{dt}}
\newcommand{\dWSFdt}{dW_{_{\rm SF}}/dt}
\newcommand{\dWSFdtX}{\frac{dW_{_{\rm SF}}}{dt}}

\newcommand{\dWSURdtX}{\frac{d\Delta W_{_{\rm SUR}}}{dt}}
\newcommand{\DWSUR}{\Delta W_{_{\rm SUR}}(t)}
\newcommand{\dxiIFdtau}{d\xi_{_{\rm IF}}/d\tau}
\newcommand{\dxiIFdtauX}{\frac{d\xi_{_{\rm IF}}}{d\tau}}
\newcommand{\dxiSUR}{\Delta\xi_{_{\rm SUR}}(\tau)}
\newcommand{\dxiSURO}{\Delta\xi_{_{\rm SUR}}(0)}

\newcommand{\dxiSURdtauX}{\frac{d\Delta\xi_{_{\rm SUR}}}{d\tau}}

\newcommand{\LIFinvsqd}{L_{_{\rm IF}}^{-2}(t)}
\newcommand{\LIF}{L_{_{\rm IF}}(t)}
\newcommand{\mH}{m_{_{\rm H}}}
\newcommand{\mbarHH}{\bar{m}_{_{\rm H_2}}}
\newcommand{\MII}{M_{_{\rm II}}(t)}
\newcommand{\MSUR}{M_{_{\rm SUR}}(t)}
\newcommand{\MSun}{{\rm M}_{_\odot}}
\newcommand{\nLAY}{n_{_{\rm LAYER}}}
\newcommand{\nIF}{n_{_{\rm IF}}(t)}
\newcommand{\nIFsqd}{n_{_{\rm IF}}^2(t)}
\newcommand{\nII}{n_{_{\rm II}}(r,t)}
\newcommand{\nIIsqd}{n_{_{\rm II}}^2(r,t)}
\newcommand{\nOO}{n_{_4}}
\newcommand{\nSUR}{n_{_{\rm SUR}}(t)}
\newcommand{\NLyC}{\dot{\cal N}_{_{\rm LyC}}}
\newcommand{\NOO}{\dot{\cal N}_{_{49}}}

\newcommand{\PIF}{P_{_{\rm IF}}(t)}
\newcommand{\Pmag}{P_{_{\rm MAG}}}
\newcommand{\RR}{{\cal R}(r,t)}
\newcommand{\SigCrit}{\Sigma_{_{\rm CRIT}}}
\newcommand{\SigLAY}{\Sigma_{_{\rm LAYER}}}
\newcommand{\tOs}{{\sc the O star}}
\newcommand{\tOss}{{\sc the O star }}
\newcommand{\RStr}{R_{_{\rm STROEMGREN}}}

\newcommand{\TII}{T_{_{\rm II}}}
\newcommand{\vSUR}{\upsilon_{_{\rm SUR}}(t)}
\newcommand{\WIF}{W_{_{\rm IF}}(t)}
\newcommand{\WIFB}{W_{_{\rm IF}}}
\newcommand{\WIFF}{W_{_{\rm IF}}(0)}
\newcommand{\WIFsqd}{W_{_{\rm IF}}^2(t)}
\newcommand{\WIFinv}{W_{_{\rm IF}}^{-1}(t)}
\newcommand{\WSF}{W_{_{\rm SF}}(t)}
\newcommand{\xiIF}{\xi_{_{\rm IF}}(\tau)}
\newcommand{\xiIFF}{\xi_{_{\rm IF}}(0)}
\newcommand{\xiIFsqd}{\xi_{_{\rm IF}}^2(\tau)}
\newcommand{\xiIFsqrt}{\xi_{_{\rm IF}}^{1/2}(\tau)}
\newcommand{\xiSF}{\xi_{_{\rm SF}}(\tau)}
\newcommand{\zIF}{z_{_{\rm IF}}}
\newcommand{\zSF}{z_{_{\rm SF}}}
\newcommand{\ZLAY}{Z_{_{\rm LAYER}}}
\newcommand{\ZLAYsqd}{Z_{_{\rm LAYER}}^2}
\newcommand{\ZOO}{Z_{_{0.1}}}

\begin{document} 

\title[Feedback from an O star in a layer]{Ionising feedback from an O star formed in a shock-compressed layer}
\author[A. P. Whitworth, F. D. Priestley \& S. T. Geen]{A. P. Whitworth$^{1}$\thanks{E-mail: ant@astro.cf.ac.uk},
F. D. Priestley$^{1}$, S. T. Geen$^{2,3}$\\
$^{1}$School of Physics and Astronomy, Cardiff University, Cardiff CF24 3AA, UK\\
$^{2}$Anton Pannekoek Institute for Astronomy, Universiteit van Amsterdam, Science Park 904, 1098 XH Amsterdam, The Netherlands\\
$^{3}$Leiden Observatory, Leiden University, PO Box 9513, 2300 RA Leiden, Netherlands}
\date{Accepted XXX. Received YYY; in original form ZZZ}
\pubyear{2019}
\label{firstpage}
\pagerange{\pageref{firstpage}--\pageref{lastpage}}
\maketitle

\begin{abstract}
We develop a simple analytic model for what happens when an O star (or compact cluster of OB stars) forms in a shock compressed layer and carves out an approximately circular hole in the layer, at the waist of a bipolar H{\sc ii} Region (H{\sc ii}R). The model is characterised by three parameters: the half-thickness of the undisturbed layer, $\ZLAY$, the mean number-density of hydrogen molecules in the undisturbed layer, $\nLAY$, and the (collective) ionising output of the star(s), $\NLyC$. The radius of the circular hole is given by $\WIF\sim 3.8\,{\rm pc}\,[\ZLAY/0.1{\rm pc}]^{-1/6}[\nLAY/10^4{\rm cm^{-3}}]^{-1/3}[\NLyC/10^{49}{\rm s^{-1}}]^{1/6}[t/{\rm Myr}]^{2/3}$. Similar power-law expressions are obtained for the rate at which ionised gas is fed into the bipolar lobes, the rate at which molecular gas is swept up into a dense ring by the shock front (SF) that precedes the ionisation front (IF), and the density in this dense ring. We suggest that our model might be a useful zeroth-order representation of many observed H{\sc ii}Rs. From viewing directions close to the midplane of the layer, the H{\sc ii}R will appear bipolar. From viewing directions approximately normal to the layer it will appear to be a limb-brightened shell but too faint through the centre to be a spherically symmetric bubble. From intermediate viewing angles more complicated morphologies can be expected.
\end{abstract}



\section{Introduction}

In \citet{WhitworthPriestley2021} we have derived an approximate analytic solution for the evolution of an H{\sc ii}R excited by an O star or compact group of OB stars (hereafter simply `\tOs') that has formed from a filament. This solution is applicable to the situation where \tOss has formed at the end of a filament, due to the gravitational focussing that occurs there \citep[e.g.][]{PonAetal2011, PonAetal2012, ClarkeWhitworth2015}, and also to the situation where \tOss has formed at the confluence of a hub-and-spoke system of filaments \citep[e.g.][]{PerettoNetal2013, PerettoNetal2014, AndersonMetal2021}, since each spoke-filament will be eroded by the ionising radiation from \tOss more-or-less independently of the other spoke-filaments.

Here we consider the situation where \tOss (again representing the case of a single O star or a compact cluster of OB stars) has formed near the centre of a shock-compressed molecular layer. For simplicity, we have in mind an approximately plane-parallel layer formed by the collision of two clouds or two converging flows \citep[e.g.][]{ElmegreenLada1977, ElmegreeBGDM1978, ScovilleNetal1986, ElmegreenB1994, Whitetal1994a, Whitetal1994b, TanJ2000, HeitschFetal2006, AuditEHennebelleP2005, HennebellePAuditE2007, HennebellePetal2007, HeitschFetal2008, DaleJetal2009, WunschRetal2010, DaleJetal2011, InoueTFukuiY2013, ChenCOstrikerE2014, TakahiraKetal2014, FukuiYetal2014, ChenCOstrikerE2015, BalfourSetal2015, DinnbierFetal2017, BalfourSetal2017, WuBetal2017a, WuBetal2017b, KohnoMetal2018, InoueTetal2018, TakahiraKetal2018, ShimaKetal2018, DobashiKetal2019, DobbsCetal2020, LiowKDobbsC2020, TanvirTDaleJ2020, DobbsCWursterJ2021, HayashiKetal2021, EnokiyaRetal2021, SakreNetal2021}.

Under this circumstance, the H{\sc ii}R excited by \tOss expands most rapidly in directions approximately normal to the layer. In these directions the IF may become density-bounded, so that ionising photons escape into the general interstellar medium. In directions close to the plane of the layer expansion of the H{\sc ii}R proceeds more slowly, and the IF is preceded by an SF which sweeps up the molecular gas in the layer into a dense ring (hereafter the swept-up ring, SUR); this SUR may become sufficiently dense and massive to spawn a second generation of stars, i.e. a variant on the collect-and-collapse mechanism for propagating star formation \citep[e.g.][]{Whitetal1994a,  DeharvengLetal2003, DaleJetal2007, ThompsonMAetal2012, PalmeirimPetal2017}. The ionised gas dispersing normal to, and on both sides of, the layer forms a Bipolar H{\sc ii}R. The inside surface of the SUR appears as a bright rim and defines the waist of the Bipolar H{\sc ii}R \citep[e.g.][]{DeharvengLetal2015, SamalMetal2018, WhitworthAetal2018}. Similar morphologies can be produced by mechanical feedback from winds and supernovae in layers \citep[e.g.][]{WareingCetal2017}, and somewhat different configurations have been proposed for more violent cloud/cloud collisions \citep[e.g.][]{OhamaAetal2018}.

\begin{figure*}
\vspace*{-4.0cm}\hspace*{-3.75cm}\includegraphics[angle=-90.,width=2.99\columnwidth]{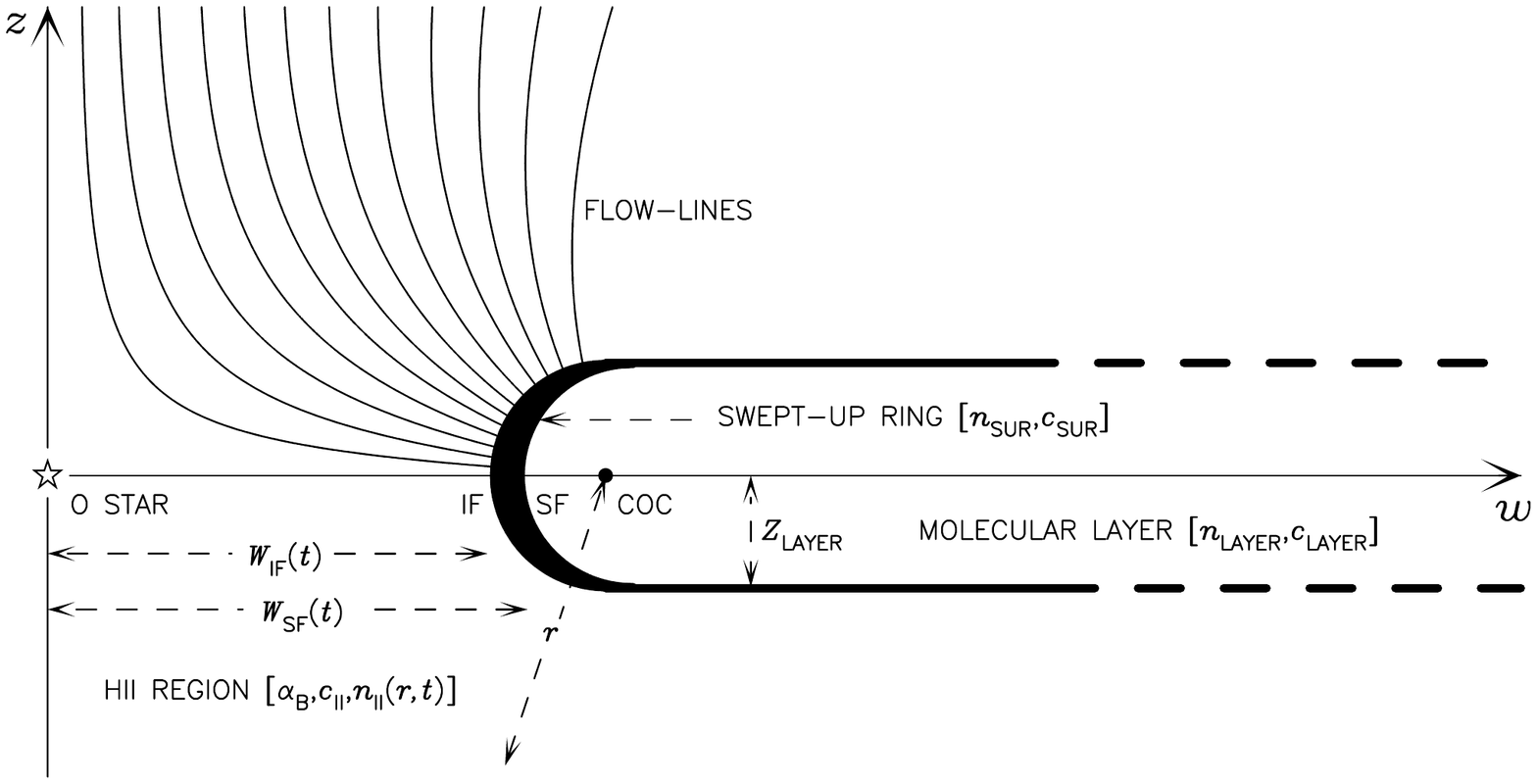}
\vspace*{-4.5cm}
\caption{Cartoon illustrating the geometry of the model on the $[w,z]$ plane. The configuration has cylindrical symmetry about the $z$ axis and reflection symmetry about the $z\!=\!0$ plane; $w\!=\![x^2\!+\!y^2]^{1/2}$. {\sc The O star} sits at the origin $[w,z]\!=\![0,0]$. The bold outline on the right represents the boundary of the `{\sc molecular layer}', which is characterised by half-thickness, $\ZLAY$, molecular-hydrogen density, $\nLAY$ and effective sound-speed, $\cLAY$. In the direction towards \tOss the molecular gas of the layer terminates in a semi-circular IF, at distance $\WIF$ from \tOs. As the IF advances into the layer, it is preceded by a SF, at distance $\WSF$ from \tOs. Between the IF and the SF is a dense ring of swept-up molecular gas, represented by the thick black meniscus-shape, and labelled `{\sc swept-up ring}' (SUR), with molecular hydrogen density $\nSUR$ and effective sound-speed $\cSUR$. The H{\sc ii}R between the \tOss and the IF is characterised by recombination coefficient, $\alfBO$, adiabatic sound-speed, $\cII$ and proton density, $\nII$. Here $r$ is distance from the centre of curvature (COC) of the IF. The ionised gas flowing off the IF disperses as indicated by the curved `{\sc flow-lines}', to produce a Bipolar H{\sc ii}R, with the {\sc swept-up ring} defining its waist.}
\label{FIG:CartoonFiL}
\end{figure*}

Viewed from directions close to the plane of the layer, both lobes of the Bipolar H{\sc ii}R may be visible. Viewed from directions approximately normal to the layer, the bright rim will appear as a ring of enhanced emission. The standard interpretation of such rings of enhanced emission is that they represent the limb-brightened projection of an approximately spherical H{\sc ii}R \citep[e.g.][]{DeharvengLetal2010, AndersonLetal2011, LiCetal2020}. However, there are some cases where the weak emission measure through the centre of the ring, and the low column-density of molecular gas, are more compatible with a Bipolar H{\sc ii}R breaking out of a layer \citep[e.g.][]{BeaumontCWilliamsJ2010, SamalMetal2018, XuJ-Letal2017, KabanovicSetal2022}. One  example of an H{\sc ii}R that can be interpreted in this way is RCW 120 \citep[e.g.][]{BeaumontCWilliamsJ2010, AndersonLetal2015, FigueiraMetal2017}, although alternative physical explanations have been proposed for this source, involving ionising stars in turbulent clouds  \citep{WalchSetal2015}, and stellar-wind bubbles \citep{MackeyJetal2016} or H{\sc ii} bubbles \citep{MarshKWhitworthA2019} due to stars that are moving supersonically relative to the ambient molecular gas. Other examples of partial or total ring-like morphologies include the shell round $\lambda$ Orionis \citep[][]{DukhangLetal2015} and G35.673-00.847 \citep[][]{DewanganLetal2018}.

The plan of the paper is as follows. In Section \ref{SEC:Model} we define a simple geometric model for the configuration. We introduce the three configuration parameters ($\ZLAY,\nLAY,\NLyC$) that define a particular realisation of the model, and the four {\it Ans{\"a}tze} that allow us to simplify the dynamical equations. In Section \ref{SEC:IonisationBalance} we formulate the equation of ionisation balance in the H{\sc ii}R, between \tOs, and the IF on the inside of the SUR. In Section \ref{SEC:Structure} we derive approximate expressions for the density and sound-speed in the SUR, and the bulk velocity of the SUR. In Section \ref{SEC:Advance} we obtain a differential equation for the position of the IF, a numerical solution to this equation, and an approximate analytic solution. In Section \ref{SEC:Dynamics} we obtain an approximate analytic expression for the rate at which ionised gas boils off the IF and escapes into the bipolar lobes. In Section \ref{SEC:Growth} we obtain a differential equation for the growth of the SUR, a numerical solution to this equation, and an approximate analytic solution. In Section \ref{SEC:Assumptions} we review the assumptions and approximations made, and attempt to identify those that might alter the results significantly. In Section \ref{SEC:Discussion} we discuss the results, and in Section \ref{SEC:Conclusions} we summarise the main results.

The essential difference between the situation considered here (a ring-shaped IF eroding a circular hole in a plane-parallel layer), and the situation considered in the earlier paper \citep[][an hemispherical IF eroding the end of a cylindrical filament]{WhitworthPriestley2021} is the geometry. Whereas the filament presents a rather small solid angle to \tOs, and one which decreases rather fast (as $\LIFinvsqd$, where $\LIF$ is the distance from \tOss to the IF where it eats into the end of the filament), the layer presents a larger solid angle to \tOs, and one which decreases more slowly (as $\WIFinv$, where $\WIF$ is the distance from \tOss to the IF where it eats into the layer). Consequently, at late times \tOss releases much more ionised gas from the layer, and sweeps up much more neutral gas into the SCR, in the situation analysed here $\;$---$\;$ than in the situation analysed by \citet{WhitworthPriestley2021}, where much less ionised gas is released from the end of the filament, and much less neutral gas is swept between the IF and the SF.

\begin{table}
\begin{center}
\caption{Definitions of acronyms and mathematical symbols.}
\begin{tabular}{lll}\hline
\multicolumn{3}{l}{\sc Acronyms} \\
H{\sc ii} Region & H{\sc ii}R & \\
Ionisation Front & IF & \\
Centre Of Curvature of IF & COC & \\
Shock Front & SF & \\
Swept-Up Ring & SUR & \\\hline
\multicolumn{3}{l}{\sc Configuration Parameters and Fiducial Values} \\
Half-thickness of layer & $\ZLAY$ & $\ZOO\times\left[0.1\,{\rm pc}\right]$ \\
Density of H$_2$ in layer & $\nLAY$ & $\nOO\times\left[10^4\,\rm{cm}^{-3}\right]$ \\
Output of ionising photons & $\NLyC$ & $\NOO\,\times\left[10^{49}\,\rm{s}^{-1}\right]$ \\
Effective sound speed in layer & $\cLAY$ & $0.31\,{\rm km\,s^{-1}}\ZOO\nOO^{1/2}$ \\
Effective sound speed in SUR & $\cSUR$ & $0.44\,{\rm km\,s^{-1}}\ZOO\nOO^{1/2}$ \\\hline
\multicolumn{3}{l}{\sc Fixed Parameters and Values} \\
Temperature in ionised gas & $\TII$ & $10^4\,{\rm K}$ \\
Sound speed in ionised gas & $\cII$ & $15\,{\rm km\,s^{-1}}$ \\
Case B recombination coeff. & $\alfBO$ & $2.7\!\times\!10^{-13}\rm{cm^3s^{-1}}$ \\\hline
\multicolumn{3}{l}{\sc Dependent Variables} \\
Distance from O star to IF & $\WIF$  & \\
Distance from O star to SF & $\WSF$ & \\
Thickness of SUR & $\DWSUR$ & \\
Density of protons at IF & $\nIF$ & \\
Density of protons in H{\sc ii}R & $\nII$ & \\
Density of H$_2$ in SUR & $\nSUR$ & \\
Bulk velocity in SUR & $\vSUR$ & \\
Mass of SUR & $\MSUR$ & \\\hline
\multicolumn{3}{l}{\sc Independent Variables} \\
Time since O star switch-on \;\;& $t$ & \\
Radial distance from COC & $r$ & \\
Distance along symmetry axis & $z$ & \\
Distance from symmetry axis & $w$ & $[x^2+y^2]^{1/2}$ \\\hline
\multicolumn{3}{l}{\sc Dimensionless Variables} \\
Dimensionless $t$ & $\tau$ & $\cLAY t/\ZLAY$ \\
Dimensionless $\WIF$ & $\xiIF$ & $\WIF/\ZLAY$ \\
Dimensionless $\WSF$ & $\xiSF$ & $\WSF/\ZLAY$ \\
Dimensionless $\DWSUR$ & $\delSUR$ & $\DWSUR/\ZLAY$ \\\hline
\end{tabular}
\end{center}
\label{TAB:Params}
\end{table}

\section{Model}\label{SEC:Model}

In this section we define the geometry of the model, introduce the three dimensionless parameters that specify a particular configuration ($\ZOO,\nOO,\NOO$), and introduce the approximations and assumptions that underlie the subsequent analysis. Most of these approximations and assumptions are evaluated retrospectively in Section \ref{SEC:Assumptions}.

We model the undisturbed layer as a semi-infinite homogeneous slab, with reflection symmetry about the $z\!=\!0$ plane (see Figure \ref{FIG:CartoonFiL}). The layer has half-thickness $\ZLAY$, i.e. it is confined to $|z|<\ZLAY$, and we define a dimensionless half-thickness
\begin{eqnarray}\label{EQN:ZOO}
\ZOO&=&\ZLAY/[0.1\,{\rm pc}].
\end{eqnarray} 

The gas has solar composition [$X=0.70, Y=0.28, Z=0.02$], and we assume that in the layer all the hydrogen is molecular. The mean mass associated with each hydrogen molecule, when contributions from all other elements, in particular helium, have been taken into account, is therefore $\mbarHH=2\mH/X=4.75\times 10^{-24}\,{\rm g}$, where $\mH$ is the mass of an hydrogen atom.

In the undisturbed layer, the number-density of molecular hydrogen is $\nLAY$, and we define a dimensionless number-density
\begin{eqnarray}\label{EQN:nOO}
\nOO&=&\nLAY/[10^4\,{\rm cm}^{-3}].
\end{eqnarray} 
It follows that the surface-density of the layer is 
\begin{eqnarray}
\SigLAY&=&2\,\ZLAY\nLAY\mbarHH\\
&=&140\,{\rm M_{_\odot}\,pc^{-2}}\,\ZOO\,\nOO\\
&\equiv&6.2\times 10^{21}\,{\rm H_{_2}\,cm^{-2}}\,\ZOO\,\nOO.
\end{eqnarray} 
We assume that the layer is sufficiently massive that it is marginally unstable against break-up. Hence the effective sound speed in the layer (due to thermal and turbulent motions) is
\begin{eqnarray}\label{EQN:cLAY.01}
\cLAY&=&\left\{\pi\,G\,\ZLAY^2\,\nLAY\,\mbarHH\right\}^{1/2}\\
&=&0.31\,{\rm km\,s^{-1}}\,\ZOO\,\nOO^{1/2}.
\end{eqnarray} 

We also assume that \tOss has formed at the centre of the layer, $[x,y,z]=[0,0,0]$, and does not move (we review this assumption in Section \ref{SEC:Movement}). At time $t=0$ \tOss starts to emit ionising photons at a constant rate $\NLyC$ (we review this assumption in Section \ref{SEC:FireUp}), and we define a dimensionless ionising output,
\begin{eqnarray}\label{EQN:NOO}
\NOO&=&\NLyC/[10^{49}\,{\rm s^{-1}}].
\end{eqnarray} 
A circular IF (implicitly an hydrogen-ionisation front) propagates outwards from \tOs, creating a circular hole in the molecular layer, which is filled by an {\sc HiiR}.

In our model, the three dimensionless parameters, $\ZOO$, $\nOO$ and $\NOO$ (Equations \ref{EQN:ZOO}, \ref{EQN:nOO} and \ref{EQN:NOO}) are the only parameters influencing the evolution of the {\sc HiiR}. They feature in many of the subsequent equations. All the variables used in the analysis, along with the acronyms, are summarised in Table 1.

The hot ionised gas in the {\sc HiiR} disperses due to both the momentum with which this gas boils off the IF, and its high thermal pressure. We assume that the ionised gas has uniform and constant gas-kinetic temperature, $\TII =10^4\,{\rm K}$, hence uniform and constant adiabatic sound speed, $\cII =15\,{\rm km\,s^{-1}}$ and uniform and constant Case B recombination coefficient, $\alfBO =2.7\times 10^{-13}\,{\rm cm^3\,s^{-1}}$.\footnote{Case B invokes the On-The-Spot Approximation, i.e. recombinations straight into the ground state of hydrogen are ignored on the assumption that such recombinations usually result in the emission of photons just above the Lyman Continuum Limit. The cross-section presented to such photons by hydrogen atoms is large. Consequently these photons are unlikely to travel far before they are re-absorbed, producing a compensatory ionisation.} We neglect the ionisation of helium. The assumption of uniform and constant temperature, and the neglect of helium ionisation, are reasonable because we shall be mainly concerned with the ionised gas near the IF. Here the ionising radiation is rather hard, and likely to maintain a high gas-kinetic temperature, $\TII \sim 10^4\,{\rm K}$. Moreover, unless the mean effective temperature of \tOss is exceptionally high, no helium-ionising photons will reach that far, i.e. the helium-ionisation fronts will both be closer to \tOs.

As the IF propagates outwards into the layer, it is preceded by a SF, which sweeps up the neutral gas of the layer. The passage of the SF will probably not change the gas-kinetic temperature much, but it is likely to amplify the turbulent motions, so that the effective sound speed in the SUR is increased somewhat. As in \citet{WhitworthPriestley2021}, we assume that $\cSUR \approx 2^{1/2}\cLAY$ (see Equations \ref{EQN:cLAY.01} and \ref{EQN:cSUR.01}); in other words, the amount of turbulent energy is approximately doubled on passing through the shock. This assumption cannot be justified precisely, but appears to be a good approximation to the amplification of turbulence behind accretion shocks, and does not affect the results significantly.\footnote{\citet{WhitJaff2018} show that molecular-line cooling effectively reduces post-shock velocities to transonic values, but the cooling then stalls due to saturation of the cooling lines. If we assume that this factor also applies to the non-thermal velocity dispersion, then the post-shock velocity dispersion should be roughly twice the pre-shock value.}

The model is concerned with (i) the dynamics of the dispersing ionised gas, (ii) the consequent advance of the IF, and (iii) the sweeping up of neutral gas immediately ahead of the IF. At the heart of the model are the following four {\it Ans{\"a}tze}.\\
{\it Ansatz 1.} The {\sc HiiR} and the SUR have reflection symmetry about the $z=0$ plane, and cylindrical symmetry about the $z$ axis (see Figure \ref{FIG:CartoonFiL}). Hence we can switch to cylindrical polar coordinates $[w,\phi,z]$, so that all quantities only depend on $w=[x^2+y^2]^{1/2}$ and $|z|$, and are independent of $\phi\!=\!\tan^{-1}(y/x)$. For brevity, we will refer to the circle defined by $[w,0\!\leq\!\phi\!<\!2\pi,z]$ as `a point' with coordinates $[w,z]$. \\
{\it Ansatz 2.} Ionised gas flowing radially off the IF diverges, and so the recombination rate decreases with distance from the IF. This divergence is largely determined by the curvature of the IF, which is of order $\ZLAY^{-1}$. We can then compute the position of the IF by integrating the rate of recombination along the $w$ axis from \tOss to the closest point on the IF at $[w,z]=[\WIF,0]$ (see Equation \ref{EQN:IonBal.02}).\\
{\it Ansatz 3.} Ionised gas flows off each point on the IF at the same rate, and this rate can be approximated by the rate at $[w,z]=[\WIF,0]$ (computed as described in {\it Ansatz 2} and Section \ref{SEC:IonisationBalance}). Evidently this overestimates the flow-rate off all other points on the IF, by a factor that increases with $|z|$, in particular because ionising radiation arrives at these points at an increasingly oblique angle to the IF as we move away from the $z\!=\!0$ plane. We compensate for this by assuming that the area of the IF is only $\sim\!4\pi[\WIF\!+\!\ZLAY]\ZLAY$ (the projected cross-sectional area of the IF as seen from \tOs) rather than the full $\sim\!2\pi^2[\WIF\!+\!\ZLAY]\ZLAY$ (the area of the convex inner surface of the SUR).\\
{\it Ansatz 4.} If the gas were to flow radially off the IF to infinity, it would converge on the $z$ axis to produce a singularity in the volume-density. We neglect this convergence, since radiation pressure and/or winds from \tOss deflect the ionised gas outwards and away from \tOs. We only take account of the divergence in the vicinity of the IF (see Figure \ref{FIG:CartoonFiL}), which is where most of the ionising radiation emitted in these directions is absorbed -- as evidenced by the brightness of `bright rims'.

With these {\it Ans{\"a}tze}, the location of the IF is given by
\begin{eqnarray}
\begin{array}{rcl}
\zIF(w,t)\!\!&\!\!=\!\!&\!\!\pm\,\left\{\ZLAY^2-[\WIF+\ZLAY-w]^2\right\}^{1/2},\\
\WIF\!\!&\!\!\leq\!\!&\!\!w\,<\,\WIF+[1\!-\!\cos(1)]\ZLAY\,,
\end{array}
\end{eqnarray} 
where $\cos(1)=0.5403\,$.

The point on the SF closest to \tOss is at $[w,z]\!=\![\WSF,0]$. Hence the thickness of the SUR, measured parallel to the $w$ axis, is
\begin{eqnarray}\label{EQN:SURWidth.01}
\DWSUR&\approx&\WSF-\WIF,
\end{eqnarray}
and the locus of the shock is
\begin{eqnarray} 
\zSF(w,t)&\approx&\zIF\!\left(w\!-\!\DWSUR,t\right).
\end{eqnarray} 
In general, $\DWSUR\ll\WIF$; in other words the SUR is quite thin, radially. This is demonstrated retrospectively in Section \ref{SEC:ThinRing}. Hence $\WSF\sim\WIF$.

Our model does not involve a magnetic field. We speculate on the consequences of including a magnetic field in Section \ref{SEC:MagneticField}.

\section{Ionisation balance}\label{SEC:IonisationBalance}

In this section we formulate the condition of ionisation balance in the space between \tOss and the IF.

For the sake of simplicity, we exploit the fact that there is an extremely short period when the gas has not had time to move far, and many of the ionising photons are being expended ionising gas for the first time. This period is very short, of order a few (say ten) recombination times,
\begin{eqnarray}
0&<&t\;\;\;\;\la\;\;\;\;\frac{5}{\alfB\,\nLAY}\;\;\;\;\sim\;\;\;\;0.00006\,{\rm Myr}\,\nOO^{-1}.\hspace{0.5cm}
\end{eqnarray} 
After this there is a period during which the IF is R-type \citep{KahnFD1954}, but this too is relatively short-lived. Thereafter, the IF switches to being D-critical \citep{KahnFD1954}, and most of the ionising radiation is expended maintaining ionisation against recombination in the region between \tOss and the IF. Only a small fraction of the ionising radiation gets to ionise new material at the advancing IF (we check this retrospectively in Section \ref{SEC:RecBal}). The newly ionised gas is significantly over-pressured, and therefore rapidly expands away from the IF.

Ionised gas flows off the D-critical IF at the adiabatic sound speed, $\cII$, and we assume that the volume-density in the ionised gas decreases approximately as $[r/\ZLAY]^{-1}$, where $r$ is distance from the centre of curvature (COC) of the IF,
\begin{eqnarray}
r&=&\left\{[\WIF+\ZLAY-w]^2+z^2\right\}^{1/2},
\end{eqnarray}
and the COC is at $[w,z]\!=\![\WIF\!+\!\ZLAY,0]\,$ (see Figure \ref{FIG:CartoonFiL}). In other words we neglect acceleration of the flow of ionised gas, as it flows away from the IF. This will reduce the recombination rate further, but will not greatly affect the advance of the IF, because the attenuation of the ionising flux is dominated by the dense region near the IF, as evident in Equation \ref{EQN:IonBal.01}. 

Since we are assuming that the \tOss is not sufficiently hot to ionise helium all the way to the IF, the volume-density of protons in the vicinity of the IF is the same as the volume-density of electrons, and we approximate this by
\begin{eqnarray}\label{EQN:nII.03}
\nII&=&\nIF\,[r/\ZLAY]^{-1},
\end{eqnarray} 
i.e. we separate the time dependence from the radial dependence. This is an acceptable approximation, provided that the timescale on which the IF advances is much longer that the timescale on which ionised gas streams far away from the IF. We show that this is indeed the case in Section \ref{SEC:SepVar}.

The recombination rate per unit volume is therefore
\begin{eqnarray}\label{EQN:RecRate.01}
\RR&=&\alfBO\,\nIIsqd\;\;\;\approx\;\;\;\alfBO\,\nIFsqd\,[r/\ZLAY]^{-2}.\hspace{0.3cm}
\end{eqnarray} 
Here, $\nIF$ is the volume-density of protons immediately outside the IF. In Section \ref{SEC:Dynamics} we obtain a closed expression for $\nIF$ (Equation \ref{EQN:nIF.01}).

On the $w$ axis ($z\!=\!0$), and in the {\sc HiiR} ($w\!<\!\WIF$), we have $r=\WIF\!+\!\ZLAY\!-\!w$. Therefore, if we (a) neglect the small fraction of ionising photons that reaches the IF and ionises new material, and (b) equate the supply of ionising photons to the rate of recombination integrated along the $w$ axis from \tOss to the IF, ionisation balance requires
\begin{eqnarray}\label{EQN:IonBal.01}
\NLyC\!\!&\!\!\!\approx\!\!\!&\!\!\int\limits_{w=0}^{w=\WIF}\;\frac{\alfBO\,\nIFsqd\,\ZLAYsqd\,4\pi w^2\,dw}{[\WIF\!+\!\ZLAY\!-\!w]^2}
\hspace{0.5cm}\\\nonumber
\!\!&\!\!\!\approx\!\!\!&\!\!4\pi\,\alfBO\,\nIFsqd\,\ZLAYsqd\left\{\frac{\left[\WIF\!+\!2\ZLAY\right]]\WIF}{\ZLAY}\right.\\\label{EQN:IonBal.02}
&&\hspace{1.0cm}\left.-2\left[\WIF\!+\!\ZLAY\right]\!\ln\!\left(\frac{\WIF\!+\!\ZLAY}{\ZLAY}\right)\right\}\!.\hspace{0.7cm}
\end{eqnarray} 
This is a key equation relating the two time-dependent quantities, $\nIF$ and $\WIF$, to the configuration parameters, $\NLyC$ and $\ZLAY$.

\section{Structure of the shock-compressed layer}\label{SEC:Structure}

The volume-density of molecular hydrogen in the SUR is
\begin{eqnarray}\label{EQN:ShockCompression.01}
\nSUR&\approx&\nLAY\left[\frac{\dWSFdt}{\cSUR}\right]^2,
\end{eqnarray} 
due to compression where the undisturbed gas in the layer is swept up into the SUR, and conservation of mass across the SF requires
\begin{eqnarray}\label{EQN:MassConservation.01}
\nLAY\,\dWSFdt&\approx&\nSUR\left[\dWSFdt - \vSUR\right]
\end{eqnarray} 
where $\vSUR$ is the velocity of the shock-compressed gas, parallel to 
the $w$ axis. Eliminating $\nSUR$ between Equations \ref{EQN:ShockCompression.01} and \ref{EQN:MassConservation.01}, we obtain
\begin{eqnarray}\label{EQN:vSUR.01}
\vSUR&\approx&\dWSFdt-\frac{\cSURsqd}{\dWSFdt}
\end{eqnarray} 

As shown in the Appendix to \citet{WhitworthPriestley2021}, gas flows into the D-critical IF at speed $\sim\cSURsqd/2\cII$ and hence
\begin{eqnarray}\label{EQN:velintoIF}
\dWIFdt-\vSUR&\approx&\cSURsqd/2\cII.
\end{eqnarray} 
Ionised gas flows off the D-critical IF at speed $\cII$, so conservation of mass across the IF gives
\begin{eqnarray}\label{EQN:ShockedDensity.01}
\nSUR&\approx&\nIF\left[\cII/\cSUR\right]^2\,.
\end{eqnarray} 
Eliminating $\nSUR$ between Equations \ref{EQN:ShockCompression.01} and \ref{EQN:ShockedDensity.01}, we obtain
\begin{eqnarray}\label{EQN:ShockSpeed.01}
\dWSFdt&\approx\cII\left[\nIF/\nLAY\right]^{1/2}
\end{eqnarray} 

If we now define the dimensionless parameter
\begin{eqnarray}\label{EQN:chi.01}
\chi(t)&=&\left[\frac{\nSUR}{\nLAY}\right]^{1/2}\;\;\,\approx\;\;\,\frac{\cII}{\cSUR}\left[\frac{\nIF}{\nLAY}\right]^{1/2}\,,
\end{eqnarray} 
Equation \ref{EQN:ShockSpeed.01} becomes
\begin{eqnarray}\label{EQN:dWSFdt.01}
\dWSFdt&\approx&\cSUR\,\chi(t)\,.
\end{eqnarray} 

Combining Equations \ref{EQN:vSUR.01} and \ref{EQN:velintoIF},
\begin{eqnarray}\label{EQN:dWIFdt.01}
\dWIFdtX\!\!&\!\!\approx\!\!&\!\!\dWSFdtX-\frac{\cSURsqd}{\dWSFdt}+\frac{\cSURsqd}{2\cII}\\\label{EQN:dWIFdt.02}
&\approx&\cSUR\left\{\chi(t)-\frac{1}{\chi(t)}\right\}\!,\hspace{0.8cm}
\end{eqnarray} 
where, to obtain the final expression (Equation \ref{EQN:dWIFdt.02}), we have dropped the third term ($\cSURsqd/2\cII$) in the expression on the preceding line  (Equation \ref{EQN:dWIFdt.01}). We justify this in Section \ref{SEC:SFSpeed}.

From Equation \ref{EQN:cLAY.01},
\begin{eqnarray}\label{EQN:cSUR.01}
\cSUR&=&\left\{2\,\pi\,G\,\ZLAY^2\,\nLAY\,\mbarHH\right\}^{1/2}\\\label{EQN:cSUR.02}
&=&0.44\,{\rm km\,s^{-1}}\,\ZOO\,\nOO^{1/2}\,;
\end{eqnarray} 
from Equations \ref{EQN:SURWidth.01}, \ref{EQN:dWSFdt.01} and \ref{EQN:dWIFdt.01},
\begin{eqnarray}\label{EQN:dWSURdt.01}
\dWSURdtX&=&\dWSFdtX-\dWIFdtX\;\;\,\approx\;\;\,\frac{\cSUR}{\chi(t)}\,;
\end{eqnarray} 
and from Equations \ref{EQN:ShockCompression.01} and \ref{EQN:dWSFdt.01},
\begin{eqnarray}\label{EQN:nSUR.01}
\nSUR&\approx&\nLAY\,\chi^2(t)\,.
\end{eqnarray} 

\section{Advance of the ionisation front}\label{SEC:Advance}

In this section we derive and solve an equation for the advance of the IF.

We first introduce the dimensionless parameter
\begin{eqnarray}\label{EQN:B.01}
K\!&\!=\!&\!\frac{\cII}{2\,\pi\,[G\mbarHH]^{1/2}\,\ZLAY^2\,\nLAY}\left[\frac{\pi\NLyC\ZLAY}{\alfBO}\right]^{1/4}\\\label{EQN:B.02}
\!&\rightarrow\!&\!34\,\ZOO^{-7/4}\,\nOO^{-1}\,\NOO^{1/4}\,,
\end{eqnarray} 
which measures the speed at which \tOss erodes the layer, and the dimensionless length and time variables,
\begin{eqnarray}
\tau&=&\frac{\cSUR t}{\ZLAY},\\
\xiIF&=&\frac{\WIF}{\ZLAY},\\
\xiSF&=&\frac{\WSF}{\ZLAY},\\
\dxiSUR&=&\frac{\DWSUR}{\ZLAY}.
\end{eqnarray} 
$\nIF$ can then be eliminated between Equations \ref{EQN:IonBal.02} and \ref{EQN:chi.01}. Equations \ref{EQN:dWIFdt.01}, \ref{EQN:cSUR.01} and \ref{EQN:B.01} are then combined to give
\begin{eqnarray}\label{EQN:dxiIFdtau.01}
\dxiIFdtauX\!\!\!&\!\!\approx\!\!&\!\!\!\chi(\tau)-\frac{1}{\chi(\tau)},\\\label{EQN:chi(tau).01}
\chi(\tau)\!\!\!&\!\!=\!\!&\!\!\!K\!\left\{\xiIFsqd+2\xiIF-2[1\!+\!\xiIF]\ln\!\left(\!1\!+\!\xiIF\right)\!\right\}^{\!-1/4}\!.\hspace{0.6cm}
\end{eqnarray} 
Equations \ref{EQN:dxiIFdtau.01} and \ref{EQN:chi(tau).01} are the equations of motion for the IF, and must be solved numerically.

We start the integration of Equation \ref{EQN:dxiIFdtau.01} with $\WIF$ equal to the Str{\o}mgren radius at the density in the layer,
\begin{eqnarray}\nonumber
\WIFF\;\;\,=\;\;\,\RStr&=&\left[\frac{3\,\NLyC}{4\,\pi\,\alfBO\,\nLAY^2}\right]^{1/3}\\
&\rightarrow&0.144\,{\rm pc}\;\nOO^{-2/3}\,\NOO^{1/3}\,,\hspace{0.5cm}
\end{eqnarray} 
hence
\begin{eqnarray}\label{EQN:xiFF0.01}
\xiIFF&\rightarrow&=1.44\,\ZOO^{-1}\,\nOO^{-2/3}\;\NOO^{1/3},
\end{eqnarray} 
but note that this choice does not affect the subsequent evolution significantly.

\begin{figure}
\vspace{-1.20cm}\hspace{-0.90cm}\includegraphics[angle=270.0,width=1.27\columnwidth]{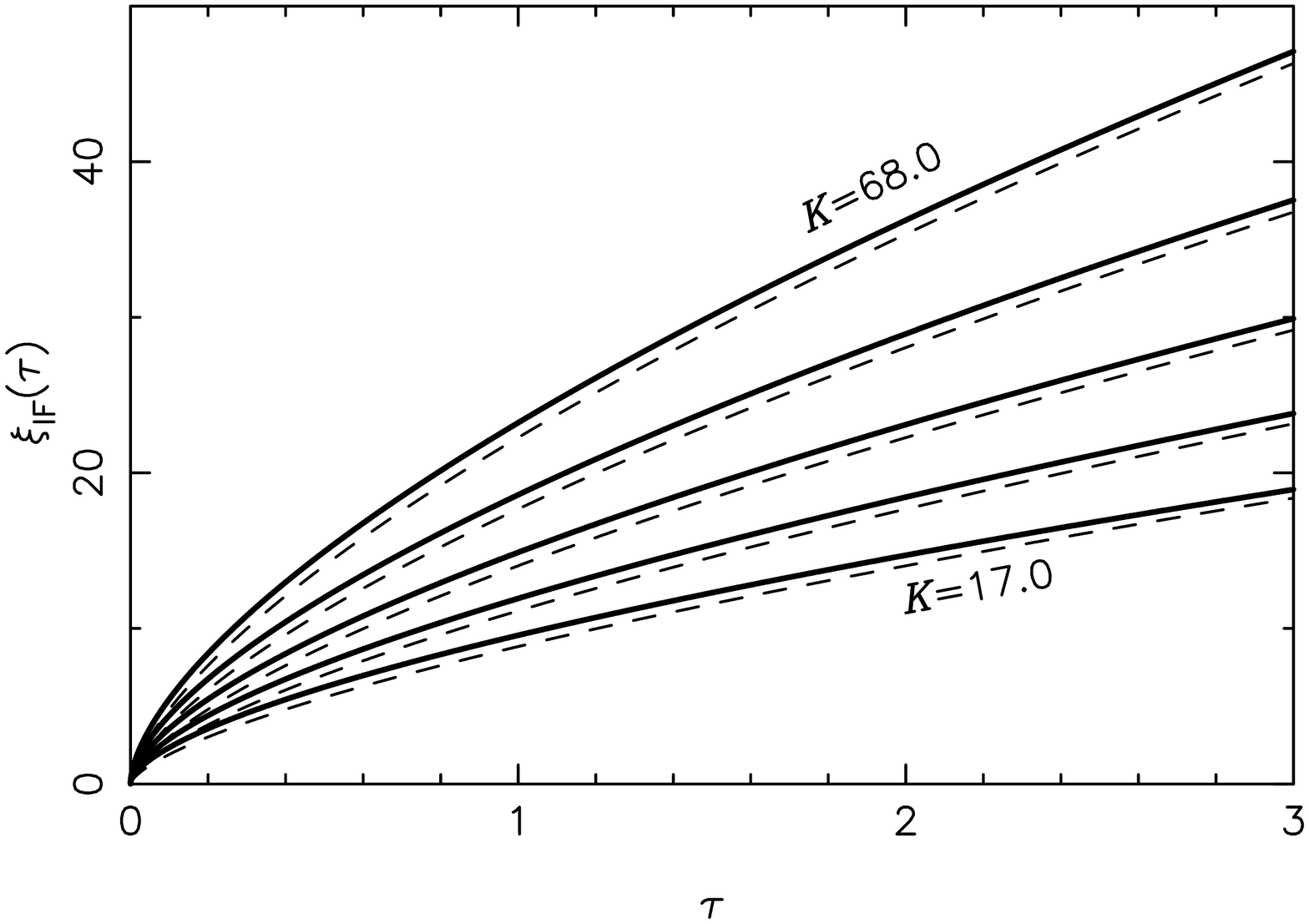}
\vspace{-0.3cm}
\caption{$\xiIF$, i.e. the distance from the \tOss to the IF, in dimensionless units, for different values of $K=17.0,\;24.0,\;34.0,\;48.1\;{\rm and}\;68.0$. The full curves show the run of $\xiIF$ obtained by numerical integration of Equations \ref{EQN:dxiIFdtau.01} and \ref{EQN:chi(tau).01}. The dashed curves show the approximate analytic solution of Equation \ref{EQN:xiIF(tau).01} for the same $K$ values.}
\label{FIG:xiIF(tau)}
\end{figure}

Accurate numerical solutions for $\xiIF$, obtained by integrating Equations \ref{EQN:dxiIFdtau.01} and \ref{EQN:chi(tau).01}, with initial conditions given by Equation \ref{EQN:xiFF0.01}, are plotted with full lines on Figure \ref{FIG:xiIF(tau)}, for five representative values of $K$, distributed about the fiducial value computed in Equation \ref{EQN:B.01}, viz. $\,17.0,\;24.0,\;34.0,\;48.1\;{\rm and}\;68.0\,$. As expected, the IF advances faster for larger values of $K$, i.e. if the layer is geometrically thinner and/or less dense, and/or the ionising output of the \tOss is larger.

We can also obtain an approximate asymptotic solution by considering the limit $\xiIF\gg1$. Equations \ref{EQN:dxiIFdtau.01} and \ref{EQN:chi(tau).01} then yield $\dxiIFdtau=K/\xiIFsqrt$ and hence
\begin{eqnarray}\label{EQN:xiIF(tau).01}
\xiIF&\sim&[3K\tau/2]^{2/3},\\\label{EQN:chi(tau).02}
\chi(\tau)&\sim&\left[2K^2/3\tau\right]^{1/3}\\
&\sim&5.5\,\ZOO^{-7/6}\,\nOO^{-5/6}\,\NOO^{1/6}\,[t/{\rm Myr}]^{-1/3}.
\end{eqnarray} 
The approximate asymptotic solution (Equation \ref{EQN:xiIF(tau).01}) is plotted with dashed lines on Figure \ref{FIG:xiIF(tau)}, for the same five values of $K$. There is very close agreement between the accurate numerical solution and the approximate asymptotic solution. Since the latter is analytic, we use it in the sequel to estimate other properties of the H{\sc ii}R and the SUR. Again we note that the advance is faster if the layer is geometrically thinner, and/or the layer is less dense, and/or the ionising output of \tOss is larger; the dependence on the ionising output is however quite weak.

Converting Equation \ref{EQN:xiIF(tau).01} back to physical variables, we obtain
\begin{eqnarray}\label{EQN:WIF.01}
\WIF&\sim&3.8\,{\rm pc}\,\ZOO^{-1/6}\,\nOO^{-1/3}\,\NOO^{1/6}\,[t/{\rm Myr}]^{2/3},\\\label{EQN:tWIF}
t(\WIFB)&\sim&0.135\,{\rm Myr}\,\ZOO^{1/4}\,\nOO^{1/2}\,\NOO^{-1/4}\,[\WIFB/{\rm pc}]^{3/2},\\\label{EQN:dWIFdt}
\dWIFdt&\sim&2.3\,{\rm km\,s^{-1}}\,\ZOO^{-1/6}\,\nOO^{-1/3}\,\NOO^{1/6}\,[t/{\rm Myr}]^{-1/3}.\hspace{0.7cm}
\end{eqnarray} 

\section{Dynamics of the overall HII region}\label{SEC:Dynamics}

In this section we solve for the rate at which ionised gas boils off the IF, and hence for the total mass of H{\sc ii} released from the filament by \tOs.

From Equation \ref{EQN:chi.01}, the density of ionised gas at the IF is
\begin{eqnarray}\label{EQN:nIF.01}
\nIF&\approx&\nLAY\left[\frac{\cSUR\,\chi(\tau)}{\cII}\right]^2\\\label{EQN:nIF.02}\label{EQN:nIF.02}
&\sim&270\,{\rm cm^{-3}}\,\ZOO^{-1/3}\,\nOO^{1/3}\,\NOO^{1/3}\,[t/{\rm Myr}]^{-2/3}.\hspace{0.7cm}
\end{eqnarray} 

Strictly speaking, the expressions and estimates derived thus far (Equations \ref{EQN:IonBal.01} through \ref{EQN:nIF.02}) pertain only to the gas on the $z$ axis. The flux of ionising radiation incident on other parts of the IF decreases with distance from the $w$ axis, largely because it arrives at increasingly oblique angles. To take account of this we invoke the third of the {\it Ans{\"a}tze} defined in Section \ref{SEC:Model}, i.e. we assume that the expressions and estimates derived thus far obtain everywhere, not just on the $w$ axis, and to compensate for this we limit the area of the IF to the cross-sectional area of the layer as seen from \tOs, $\,\approx 4\pi[\WIF\!+\!\ZLAY]\ZLAY\sim4\pi\WIF\ZLAY$.

The net rate at which ionised gas flows off the IF is therefore
\begin{eqnarray}
\dMIIdt&\sim&\pi\,\WIF\,\ZLAY\,\nIF\,\mbarHH\,\cII\\
&\sim&350\,{\rm M_{_\odot}\,Myr^{-1}}\,\ZOO^{1/2}\,\NOO^{1/2}\,,
\end{eqnarray} 
and the total mass boiled off the IF is
\begin{eqnarray}\label{EQN:MII.01}
\MII&\sim&350\,{\rm M_{_\odot}}\,\ZOO^{1/2}\,\NOO^{1/2}\,[t/{\rm Myr}]\,.
\end{eqnarray} 
As we show below (Section \ref{SEC:Growth}), this is quite a modest mass, as compared with the mass that is swept up into the SUR.

\begin{figure}
\vspace{-1.20cm}\hspace{-0.90cm}\includegraphics[angle=270.0,width=1.27\columnwidth]{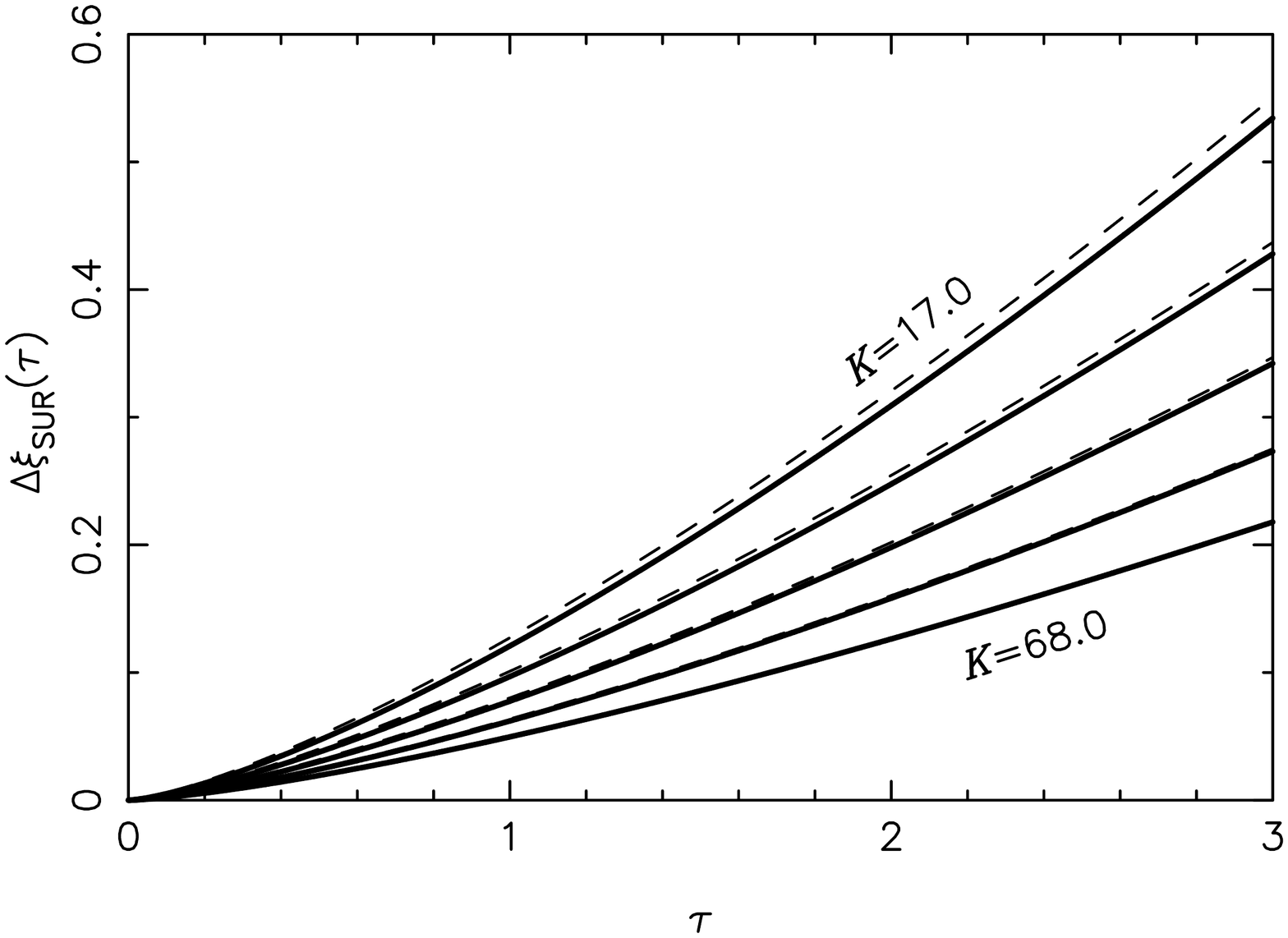}
\vspace{-0.3cm}
\caption{$\dxiSUR$, i.e. the radial thickness of the SUR, in dimensionless units, for different values of $K=17.0,\;24.0,\;34.0,\;48.1\;{\rm and}\;68.0$. The full curves show the run of $\dxiSUR$ obtained by numerical integration of Equations \ref{EQN:dxiSURdtau.01} and \ref{EQN:chi(tau).01}. The dashed curves show the approximate analytic solution of Equation \ref{EQN:dxiSUR.02} for the same $K$ values.}
\label{FIG:dxiSUR(tau)}
\end{figure}

\section{Growth of the shock-compressed ring}\label{SEC:Growth}

In this section we derive the parameters of the SUR (thickness, mass and expansion velocity).

The thickness of the SUR grows at a rate given by Equation \ref{EQN:dWSURdt.01}, and hence in terms of dimensionless variables,
\begin{eqnarray}\label{EQN:dxiSURdtau.01}
\dxiSURdtauX&\approx&\frac{1}{\chi(\tau)}
\end{eqnarray} 
Accurate numerical solutions for $\dxiSUR$ obtained by integrating Equation \ref{EQN:dxiSURdtau.01} with $\chi(\tau)$ from Equation \ref{EQN:chi(tau).02}, the initial condition $\dxiSURO=0$, and $K=17.0,\;24.0,\;34.0,\;48.1\;{\rm and}\;68.0\,$, are plotted with full lines on Figure \ref{FIG:dxiSUR(tau)}.

In the limit that $\xiIF\gg1$, we can substitute for $\chi(\tau)$ in Equation \ref{EQN:dxiSURdtau.01} from Equation \ref{EQN:chi(tau).02} to obtain
\begin{eqnarray}\label{EQN:dxiSURdtau.02}
\dxiSURdtauX&\sim&\left[\frac{3\,\tau}{2\,K^2}\right]^{1/3},
\end{eqnarray} 
and hence an approximate analytic solution,
\begin{eqnarray}\label{EQN:dxiSUR.02}
\dxiSUR&\sim&\left[81\tau^4/128K^2\right]^{1/3}
\end{eqnarray} 
The approximate asymptotic solution (Equation \ref{EQN:dxiSUR.02}) is plotted with dashed lines on Figure \ref{FIG:dxiSUR(tau)}, for the same five values of $K$. Again there is very good agreement between the accurate numerical solution and the approximate asymptotic solution, and we adopt the latter because it is analytic. Converting back to physical variables, Equation \ref{EQN:dxiSUR.02} gives
\begin{eqnarray}\label{EQN:DWSUR.02}
\DWSUR&\sim&0.06\,{\rm pc}\,\ZOO^{13/6}\,\nOO^{4/3}\,\NOO^{-1/6}\,[t/{\rm Myr}]^{4/3}.\hspace{0.6cm}
\end{eqnarray} 

The density in the SUR is given by Equation \ref{EQN:nSUR.01},
\begin{eqnarray}
\nSUR\!\!&\!\!\sim\!\!&\!\!3.1\times 10^5\,{\rm H_{_2}\,cm^{-3}}\,\ZOO^{-7/3}\,\nOO^{-2/3}\,\NOO^{1/3}\,[t/{\rm Myr}]^{-2/3}.\hspace{0.7cm}
\end{eqnarray} 

The total mass swept up by the SF is given by
\begin{eqnarray}\label{EQN:MSU.01}
\MSUR\!&\!\!\approx\!\!&\!\pi\,\WIFsqd\,\SigLAY\\\label{EQN:MSU.02}
\!&\!\!\sim\!\!&\!6400\,\MSun\,\ZOO^{2/3}\,\nOO^{1/3}\,\NOO^{1/3}\,[t/{\rm Myr}]^{4/3}.\hspace{1.3cm}
\end{eqnarray} 
Comparing this with the total mass boiled off the IF (Equation \ref{EQN:MII.01}), we see that, if we adopt the fiducial parameters, then at late times $t\gtrsim 0.1\,{\rm Myr}$, most of the swept-up mass is in the SUR rather than the H{\sc ii}R. This dense SCR is likely to spawn a second generation of stars. The swept-up mass is larger for geometrically thicker layers (larger $\ZOO$), denser layers (larger $\nOO$), and larger ionising output (larger $\NOO$).

The outward velocity of the SUR is given by
\begin{eqnarray}
\vSUR&\sim&\cSUR\chi(t)\\
&\rightarrow&2.4\,{\rm km\,s^{-1}}\ZOO^{-1/6}\,\nOO^{-1/3}\,\NOO^{1/6}\,[t/{\rm Myr}]^{-1/3}.\hspace{1.1cm}
\end{eqnarray} 

\section{Assumptions and Approximations}\label{SEC:Assumptions}

In this section we revisit the approximations and assumptions made in the preceding analysis, and show that they are appropriate.

\subsection{Movement of \tOs}\label{SEC:Movement}

We have assumed that \tOss is stationary. The predictions of the model cannot be applied if the velocity of \tOss -- in particular the component normal to the layer -- is comparable with the speed of advance of the IF (Equation \ref{EQN:dWIFdt}). Under the circumstance that \tOss is actually a compact cluster of OB stars, we must consider both the bulk velocity of the `centre of ionising output', {\it and} the velocities with which the cluster disperses. The model is only applicable to cases where these velocities are relatively small, say $\lesssim1\,{\rm km\,s^{-1}}$. Since newly formed O stars frequently have velocities exceeding this threshold, and clusters of OB stars may also have velocity dispersions exceeding the threshold, the model will not always be applicable to massive stars formed in a shock-compressed layer, only to those where the massive stars end up with relatively low velocities -- that is, excluding the systematic velocities in multiple systems and the random velocities in a tightly bound compact cluster.

\subsection{Constancy of the ionising output, $\NLyC$}\label{SEC:FireUp}

We have assumed that \tOss switches on instantaneously, and then emits ionising photons at a constant rate, $\NLyC$. Given the timescales involved in the evolution of an O star ($\gtrsim 3\,{\rm Myr}$), and in the advance of the IF (Equation \ref{EQN:tWIF}), this is a reasonable assumption for times $0.1\,{\rm Myr}\lesssim t\lesssim3\,{\rm Myr}$, and possibly longer, depending on the precise masses of the stars involved. There will be a short period when the {\sc HiiR} is small and approximately spherical, and then a short period when it has just broken out of the layer. These two periods together last less than $0.1\,{\rm Myr}$, and once they are over, the basic geometry of the model, as sketched in Figure \ref{FIG:CartoonFiL}, should be valid.

\subsection{Thickness of the SUR}\label{SEC:ThinRing}

We have assumed that $\DWSUR\ll\WIF$, i.e. that the radial thickness of the SUR is very small compared with its radius, and hence that $\WSF\simeq\WIF$. Combining Equations \ref{EQN:WIF.01} and \ref{EQN:DWSUR.02}, we obtain
\begin{eqnarray}\label{EQN:DWSUR:WIF}
\DWSUR/\WIF\!&\!\sim\!&\!0.016\,\ZOO^{7/3}\,\nOO^{5/3}\,\NOO^{-1/3}\,[t/{\rm Myr}]^{2/3}.\hspace{0.5cm}
\end{eqnarray} 
Thus, unless the thickness and/or density of the undisturbed layer are very large, and/or the erosion has been ongoing for a very long time, $\;\DWSUR\!\ll\!\WIF$, and so to a first approximation we are justified in setting $\WSF\sim\WIF$.

\subsection{Fraction of ionising photons reaching the IF}\label{SEC:RecBal}

In deriving Equation \ref{EQN:IonBal.01}, which constrains the position of the IF by requiring a balance between ionisation and recombination on the line from \tOss to the IF, we have ignored the ionising photons that reach the IF and ionise new material. Since ionised gas flows off the IF at speed $\cII$, the number-flux of hydrogen-ionising photons impinging on the IF is given by $\sim\!\nIF\,\cII$. The number-flux of ionising photons that would impinge on the IF if there were no absorption (to balance recombination) in the region between \tOss and the IF is $\NLyC/4\pi\WIFsqd$. Thus the fraction of ionising radiation expended producing new ionisations at the IF is
\begin{eqnarray}\label{EQN:fraction.01}
\frac{\nIF\cII}{\NLyC/4\pi\WIFsqd}\!\!&\!\sim\!&\!\!0.07\,\ZOO^{-2/3}\,\nOO^{-1/3}\,\NOO^{-1/3}\,[t/{\rm Myr}]^{2/3}.\hspace{0.6cm}
\end{eqnarray} 
This fraction is small and remains small ($\lesssim 0.20$) for $t\lesssim 5\,{\rm Myr}$, so we are justified in neglecting it in deriving Equation \ref{EQN:IonBal.01}.

\subsection{Separation of variables}\label{SEC:SepVar}

In deriving Equation \ref{EQN:nII.03}, we have split the $r$-dependence of $\nII$ from the $t$-dependence. This split is acceptable provided that the IF advances on a timescale much longer than the timescale it takes for newly ionised gas to get far from the IF, which reduces to the condition
\begin{eqnarray}
\frac{\WIF}{\dWIFdt}&\gg&\frac{\ZLAY}{\cII}.
\end{eqnarray} 
Substituting from Equations \ref{EQN:WIF.01} and \ref{EQN:dWIFdt}, the timescale on which the IF advances is
\begin{eqnarray}
\frac{\WIF}{\dWIFdt}&\sim&1.65\,{\rm Myr}.
\end{eqnarray}
The timescale on which ionised gas flows away from the IF is
\begin{eqnarray}\label{EQN:DispersalTime}
\frac{\ZLAY}{\cII}&\sim&0.0065\,{\rm Myr}\,\ZOO.
\end{eqnarray} 
Evidently separation of the $r$-dependence from the $t$-dependence is justified.

\subsection{Speed of SF}\label{SEC:SFSpeed}

In deriving Equation \ref{EQN:dWIFdt.02} from Equation \ref{EQN:dWIFdt.01}, we have dropped the term $\,-\,\cSUR^2/2\cII$, on the assumption that $\dWSFdt\!\ll\!2\cII$. Since $\dWSFdt\simeq\dWIFdt$, Equation \ref{EQN:dWIFdt} gives
\begin{eqnarray}\label{EQN:dWSFdt.07}
\dWSFdt&\sim&2.3\,{\rm km\,s^{-1}}\,\ZOO^{-1/6}\,\nOO^{-1/3}\,\NOO^{1/6}\,[t/{\rm Myr}]^{-1/3}.\hspace{0.6cm}
\end{eqnarray}
Evidently $\dWSFdt\ll2\cII\simeq 30\,{\rm km\,s^{-1}}$, except at very early times where the analysis presented here is not expected to apply (see discussion at the start of Section \ref{SEC:IonisationBalance}).

Having placed an upper limit on the speed of the SF, we should also check that the SF advances sufficiently fast to compress the gas it sweeps up. Comparing $\dWSFdt$ from Equation \ref{EQN:dWSFdt.07}, with $\cSUR$ from Equation \ref{EQN:cSUR.02}, we see that there will only be a shock if the layer is not too thick and not too dense, and the ionising output is not too small.

\subsection{Neglect of magnetic field}\label{SEC:MagneticField}

Since observations \citep[e.g.][]{WardThompsonDetal2017, FisselLetal2019, SoamAetal2019, PillaiTetal2020, YasuoDetal2020, ArzoumanianDetal2021, KwonWetal2022, PattleKetal2022}, and simulations \citep[e.g.][]{SolerJetal2013, SeifriedDWalchS2015, GirichidisPetal2018, GomezGetal2018} suggest that the gas flows assembling dense star forming gas structures tend to involve velocities parallel or anti-parallel to the large-scale magnetic field, we limit discussion to a large-scale magnetic field parallel to the $z$ axis. In this situation, there are two effects that should be considered. Firstly, the field will inhibit gravitational fragmentation of the layer (see Section \ref{SEC:GravStab}). Secondly, the field will resist the advance of the ionisation front (see Section \ref{SEC:MagIF}). We note that large-scale magnetic fields in the interstellar medium are typically $\lesssim 10\,\mu{\rm G}$ \citep[e.g.][]{HeilesCTrolandT2005}.

\subsubsection{Stability of the layer against gravitational fragmentation}\label{SEC:GravStab}

If the undisturbed field is $\boldsymbol{B}\!=\!\BO\hat{\boldsymbol e}_z$, the critical surface-density for lateral contraction and fragmentation of the layer is
\begin{eqnarray}
\SigCrit&\sim&\left[\frac{5}{G}\right]^{1/2}\,\frac{\BO}{3\pi}\;\;\sim\;\;44\,{\rm M_\odot\,pc^{-2}}\,\left[\frac{\BO}{10\,\mu{\rm G}}\right]\hspace{0.6cm}
\end{eqnarray}
\citep[][his equation 85]{MestelL1965b}. Thus the magnetic field will only have a significant influence on the dynamics of the undisturbed layer if $\SigLAY\lesssim\SigCrit$, i.e. if
\begin{eqnarray}
\BO&\gtrsim&32\,\mu{\rm G}\;\ZOO\,\nOO.
\end{eqnarray}

\subsubsection{Magnetic resistance to the advance of the IF}\label{SEC:MagIF}

The ram-pressure of the ionised gas flowing off the IF is
\begin{eqnarray}
\PIF\!\!&\!\!\approx\!\!&\!\!\frac{\nIF\,\mbarHH\cII^2}{2}\\
\!\!&\!\!\sim\!\!&\!\!1.4\times 10^{-9}\,{\rm erg\,cm^{-3}}\,\ZOO^{-1/3}\,\nOO^{1/3}\,\NOO^{1/3}\,[t/{\rm Myr}]^{-2/3}\!.\hspace{0.6cm}
\end{eqnarray}
The magnetic pressure is
\begin{eqnarray}
\Pmag&=&\frac{\BO^2}{8\pi}\;\;\sim\;\;4\times 10^{-12}\,{\rm erg\,cm^{-3}}\,\left[\frac{\BO}{10\,\mu{\rm G}}\right]^2.
\end{eqnarray}
Therefore expansion of the newly ionised gas in the immediate vicinity of the IF is not significantly slowed by the magnetic field. Moreover, as the gas flows further away from the IF its ram-pressure only decreases quite slowly (approximately as $r^{-1}$). Since the ionisation balance is dominated by recombinations in the immediate vicinity of the IF, the magnetic field should have little effect on the advance of the IF. In the {\sc HiiR} the magnetic field will simply be advected by the ionised gas.

\subsection{Star formation in the SUR}

It is certainly possible that a second generation of stars will form in the SUR. However, without knowing details of the turbulence there (i.e. how much is in wavelengths that can be rendered unstable against collapse), it is impossible to evaluate this possibility. Such considerations lie outside the scope of the current paper, and would almost certainly have to be addressed with numerical simulations, and a slew of additional parameters regulating the initial and boundary conditions,  and the constitutive physics.

\section{Discussion}\label{SEC:Discussion}

We have explored the possibility that, if an O star, or compact group of OB stars, forms in a shock-compressed layer {\it and} stays close to the mid-plane of the layer, then it will carve out an approximately circular hole in the layer. Ionised gas will steam away above and below the layer, producing a bipolar H{\sc ii}R. At the waist of the bipolar H{\sc ii}R, an IF will be driven into the layer, producing an approximately circular bright rim, and sweeping up a ring of dense shocked gas, within which a second generation of stars may form.

\subsection{Observed morphology}

The observed morphology of such a configuration will depend strongly on the viewing angle, relative to the normal to the layer (here the $z$ axis) and on the wavelength, due to dust extinction.

Observers who are close to $z$ axis, and therefore view the layer close to face-on, will see an approximately circular bright rim, which might in the first instance be interpreted with a limb-brightened spherically symmetric model -- except that the column-densities of ionised {\it and} neutral gas through the middle of the ring will be significantly lower than expected. Hence the model might offer a straightforward explanation for the cylindrical H{\sc ii}Rs observed by \citet{BeaumontCWilliamsJ2010}.

Observers close to the $z\!=\!0$ plane will see a bipolar H{\sc ii}R. Any difference between the size and brightness of the lobes would be attributable to the source (or sources) of ionising radiation having moved away from the midplane of the layer, although the H{\sc ii}R would cease to be bipolar if this movement were too great. Further asymmetry might also derive from differential extinction due to the dust in the layer.

It follows that there will probably be a range of intermediate angles relative to the $z$ axis, from which the projected morphology of the H{\sc ii}R  -- even if it is intrinsically bipolar, and conforms to our model -- will be quite complicated and hard to interpret.

\subsection{Overall dynamics and mass budget}

The IF advances quite rapidly into the the layer, typically at speeds exceeding $1\,{\rm km\,s^{-1}}$, sweeping up mass into a dense ring where a second generation of stars may form. Ionised gas flows off the IF and escapes above and below the layer at an approximately constant rate, but most of the mass that is shifted to create the hole in the layer ends up in  the dense ring between the IF and the SF.

\subsection{Comparison with feedback in a filament}

The expressions derived here for the rate at which the IF advances (Equation \ref{EQN:dWIFdt}), the mass of ionised gas (Equation \ref{EQN:MII.01}) and the mass of gas in the swept-up ring (Equation \ref{EQN:MSU.02}) can be compared with those obtained in \citet{WhitworthPriestley2021} for the case of an O star formed in, or at the end of, a filament.

The rate of advance of the IF has approximately the same dependence on the half-thickness of the undisturbed layer ({\it vice} the radius of the undisturbed filament), the density in the undisturbed layer ({\it vice} the density in the undisturbed filament), the ionising output of the O star and the elapsed time since switch-on. This is because the conditions at the ionisation front are largely dictated by the ionising output, $\NLyC$, and by physical conditions ahead of the IF (i.e. the density and sound speed in the undisturbed layer ({\it vice} filament).

However, the mass of ionised gas flowing off the IF and the mass of neutral gas swept up between the IF and the SF, both have different dependences, due to the different geometries. For the layer considered here, the mass of ionised gas increases linearly with time, whereas for the filament the mass of ionised gas only increases as $t^{1/3}$. Similarly, for the layer, the mass of swept up neutral gas increases as $t^{4/3}$, whereas for the filament it only increases as $t^{2/3}$. This is basically because, as the IF moves further away from the O star, the solid angle presented by the (edge-on) layer decreases much more slowly than the solid angle presented by the (end-on) filament.

We note that the advance of the IF into the layer does not stall, and nor does the advance of the IF into the filament in \citet{WhitworthPriestley2021}. If  \tOss were to emit ionising radiation for ever, and the layer were of infinite two-dimensional extent, or the filament were of infinite length, the IF would continue to advance indefinitely, albeit it at an ever decreasing rate. This is because, in these two situations, the ionised gas boiling off the IF diverges and escapes, in principle to infinity; it does not stick around for long on the path taken by ionising photons from \tOss to the IF, and therefore its re-ionisation (following recombination) does not use up many of these photons. The advance of the IF slows down because the flux of ionising radiation reaching the IF decreases due to geometric dilution. This contrasts with the situation where an {\sc O star} is placed in a uniform medium and ionises a spherical H{\sc ii}R. In that case the resulting H{\sc ii}R expands until there is pressure balance between the relatively diffuse but hot gas of the H{\sc ii}R and the relatively dense but cold gas of the surrounding neutral medium, and the advance of the IF then stalls \citep[e.g.][]{BisbasTetal2015}; thereafter, all the ionising output of the \tOss is expended balancing recombination in a static, spherical H{\sc ii}R.

\section{Conclusions}\label{SEC:Conclusions}

We have developed a simple model for the feedback from an O star (or equivalently a compact cluster of OB stars), formed in a dense layer such as might result from the collision of two clouds or two colliding streams.

Similarly to the model developed in \citet{WhitworthPriestley2021} for feedback from an O star formed in, or at the end of, a filament, the model developed here is based on a number of simplifying assumptions, in particular (i) that the O star remains close to the mid-plane of the layer, (ii) that the resulting {\sc HiiR} is approximately circular when projected onto the mid-plane of the layer; (iii) that the radius of curvature of the ionisation front, where it eats into the layer, is approximately equal to the half-thickness of the layer.

We derive power law expressions for the key parameters, in terms of $\ZOO\!=\!\ZLAY/0.1{\rm pc}$ (where $\ZLAY$ is the half-thickness of the undisturbed layer),  $\nOO\!=\!\nLAY/10^4{\rm cm^{-3}}$ (where $\nLAY$ is the volume-density of molecular hydrogen in the undisturbed layer), and $\NOO\!=\!\NLyC/10^{49}{\rm s^{-1}}$ (where $\NLyC$ is the rate at which the O star emits ionising photons).

The distance from \tOss to the IF, $\WIF$, and the speed of advance of the IF are given by
\begin{eqnarray}
\WIF\!&\!\!\sim\!\!&\!3.8\,{\rm pc}\,\ZOO^{-1/6}\,\nOO^{-1/3}\,\NOO^{1/6}\,[t/{\rm Myr}]^{2/3},\\
\dWIFdt\!&\!\!\sim\!\!&\!2.3\,{\rm km\,s^{-1}}\,\ZOO^{-1/6}\,\nOO^{-1/3}\,\NOO^{1/6}\,[t/{\rm Myr}]^{-1/3}
\end{eqnarray}
(see Section \ref{SEC:Advance}).

The rate at which ionised gas is boiled off the ionisation front is given by
\begin{eqnarray}
\dMIIdt&\sim&350\,\MSun\,{\rm Myr^{-1}}\,\ZOO^{1/2}\,\NOO^{1/2}\,,
\end{eqnarray} 
and is constant. The total mass of ionised gas is
\begin{eqnarray}
\MII&\sim&350\,\MSun\,\ZOO^{1/2}\,\NOO^{1/2}\,[t/{\rm Myr}]
\end{eqnarray}
(see Section \ref{SEC:Dynamics}). 

The rate at which the gas in the layer is swept up and compressed into a ring immediately ahead of the ionisation front is
\begin{eqnarray}\label{EQN:dMSUdt.01}
\dMSURdt\!\!&\!\!\!\sim\!\!\!&\!\!8500\,\MSun\,{\rm Myr^{-1}}\,\ZOO^{2/3}\,\nOO^{1/3}\,\NOO^{1/3}\,[t/{\rm Myr}]^{1/3},\hspace{0.6cm}
\end{eqnarray} 
and the mass of the ring is
\begin{eqnarray}
\MSUR\!&\!\!\sim\!\!&\!6400\,\MSun\,\ZOO^{2/3}\,\nOO^{1/3}\,\NOO^{1/3}\,[t/{\rm Myr}]^{4/3}\hspace{1.0cm}
\end{eqnarray} 
(see Section \ref{SEC:Growth}).

The simplifying assumptions made in order to derive these expressions are primarily contained in the {\it Ans{\"a}tze} (Section \ref{SEC:Model}). The key approximations made to justify the subsequent analysis, and hence the derivation of the above approximate expressions, are justified retrospectively in Section \ref{SEC:Assumptions}.

\section*{Acknowledgements}
APW and FDP gratefully acknowledge the support of an STFC Consolidated Grant (ST/K00926/1). STG gratefully acknowledges the support of a Spinoza award from the Dutch Science Organization (NWO) for research on the physics and chemistry of the interstellar medium. We thank the referee, Dr. Tom Haworth, for his constructive, encouraging and prompt report.

\section*{Data Availability}
All software used will be supplied on request to APW.

\bibliographystyle{mnras}
\bibliography{WhitworthReferences}

\begin{thebibliography}{}
\makeatletter
\relax
\def\mn@urlcharsother{\let\do\@makeother \do\$\do\&\do\#\do\^\do\_\do\%\do\~}
\def\mn@doi{\begingroup\mn@urlcharsother \@ifnextchar [ {\mn@doi@}
  {\mn@doi@[]}}
\def\mn@doi@[#1]#2{\def\@tempa{#1}\ifx\@tempa\@empty \href
  {http://dx.doi.org/#2} {doi:#2}\else \href {http://dx.doi.org/#2} {#1}\fi
  \endgroup}
\def\mn@eprint#1#2{\mn@eprint@#1:#2::\@nil}
\def\mn@eprint@arXiv#1{\href {http://arxiv.org/abs/#1} {{\tt arXiv:#1}}}
\def\mn@eprint@dblp#1{\href {http://dblp.uni-trier.de/rec/bibtex/#1.xml}
  {dblp:#1}}
\def\mn@eprint@#1:#2:#3:#4\@nil{\def\@tempa {#1}\def\@tempb {#2}\def\@tempc
  {#3}\ifx \@tempc \@empty \let \@tempc \@tempb \let \@tempb \@tempa \fi \ifx
  \@tempb \@empty \def\@tempb {arXiv}\fi \@ifundefined
  {mn@eprint@\@tempb}{\@tempb:\@tempc}{\expandafter \expandafter \csname
  mn@eprint@\@tempb\endcsname \expandafter{\@tempc}}}

\bibitem[\protect\citeauthoryear{{Anderson}, {Bania}, {Balser}  \&
  {Rood}}{{Anderson} et~al.}{2011}]{AndersonLetal2011}
{Anderson} L.~D.,  {Bania} T.~M.,  {Balser} D.~S.,   {Rood} R.~T.,  2011,
  \mn@doi [\apjs] {10.1088/0067-0049/194/2/32}, \href
  {https://ui.adsabs.harvard.edu/abs/2011ApJS..194...32A} {194, 32}

\bibitem[\protect\citeauthoryear{{Anderson} et~al.,}{{Anderson}
  et~al.}{2015}]{AndersonLetal2015}
{Anderson} L.~D.,  et~al., 2015, \mn@doi [\apj] {10.1088/0004-637X/800/2/101},
  \href {https://ui.adsabs.harvard.edu/abs/2015ApJ...800..101A} {800, 101}

\bibitem[\protect\citeauthoryear{{Anderson} et~al.,}{{Anderson}
  et~al.}{2021}]{AndersonMetal2021}
{Anderson} M.,  et~al., 2021, \mn@doi [\mnras] {10.1093/mnras/stab2674}, \href
  {https://ui.adsabs.harvard.edu/abs/2021MNRAS.508.2964A} {508, 2964}

\bibitem[\protect\citeauthoryear{{Arzoumanian} et~al.,}{{Arzoumanian}
  et~al.}{2021}]{ArzoumanianDetal2021}
{Arzoumanian} D.,  et~al., 2021, \mn@doi [\aap] {10.1051/0004-6361/202038624},
  \href {https://ui.adsabs.harvard.edu/abs/2021A&A...647A..78A} {647, A78}

\bibitem[\protect\citeauthoryear{{Audit} \& {Hennebelle}}{{Audit} \&
  {Hennebelle}}{2005}]{AuditEHennebelleP2005}
{Audit} E.,  {Hennebelle} P.,  2005, \mn@doi [\aap]
  {10.1051/0004-6361:20041474}, \href
  {https://ui.adsabs.harvard.edu/abs/2005A&A...433....1A} {433, 1}

\bibitem[\protect\citeauthoryear{{Balfour}, {Whitworth}, {Hubber}  \&
  {Jaffa}}{{Balfour} et~al.}{2015}]{BalfourSetal2015}
{Balfour} S.~K.,  {Whitworth} A.~P.,  {Hubber} D.~A.,   {Jaffa} S.~E.,  2015,
  \mn@doi [\mnras] {10.1093/mnras/stv1772}, \href
  {https://ui.adsabs.harvard.edu/abs/2015MNRAS.453.2471B} {453, 2471}

\bibitem[\protect\citeauthoryear{{Balfour}, {Whitworth}  \& {Hubber}}{{Balfour}
  et~al.}{2017}]{BalfourSetal2017}
{Balfour} S.~K.,  {Whitworth} A.~P.,   {Hubber} D.~A.,  2017, \mn@doi [\mnras]
  {10.1093/mnras/stw2956}, \href
  {https://ui.adsabs.harvard.edu/abs/2017MNRAS.465.3483B} {465, 3483}

\bibitem[\protect\citeauthoryear{{Beaumont} \& {Williams}}{{Beaumont} \&
  {Williams}}{2010}]{BeaumontCWilliamsJ2010}
{Beaumont} C.~N.,  {Williams} J.~P.,  2010, \mn@doi [\apj]
  {10.1088/0004-637X/709/2/791}, \href
  {https://ui.adsabs.harvard.edu/abs/2010ApJ...709..791B} {709, 791}

\bibitem[\protect\citeauthoryear{{Bisbas} et~al.,}{{Bisbas}
  et~al.}{2015}]{BisbasTetal2015}
{Bisbas} T.~G.,  et~al., 2015, \mn@doi [\mnras] {10.1093/mnras/stv1659}, \href
  {https://ui.adsabs.harvard.edu/abs/2015MNRAS.453.1324B} {453, 1324}

\bibitem[\protect\citeauthoryear{{Chen} \& {Ostriker}}{{Chen} \&
  {Ostriker}}{2014}]{ChenCOstrikerE2014}
{Chen} C.-Y.,  {Ostriker} E.~C.,  2014, \mn@doi [\apj]
  {10.1088/0004-637X/785/1/69}, \href
  {https://ui.adsabs.harvard.edu/abs/2014ApJ...785...69C} {785, 69}

\bibitem[\protect\citeauthoryear{{Chen} \& {Ostriker}}{{Chen} \&
  {Ostriker}}{2015}]{ChenCOstrikerE2015}
{Chen} C.-Y.,  {Ostriker} E.~C.,  2015, \mn@doi [\apj]
  {10.1088/0004-637X/810/2/126}, \href
  {https://ui.adsabs.harvard.edu/abs/2015ApJ...810..126C} {810, 126}

\bibitem[\protect\citeauthoryear{{Clarke} \& {Whitworth}}{{Clarke} \&
  {Whitworth}}{2015}]{ClarkeWhitworth2015}
{Clarke} S.~D.,  {Whitworth} A.~P.,  2015, \mn@doi [\mnras]
  {10.1093/mnras/stv393}, \href
  {https://ui.adsabs.harvard.edu/abs/2015MNRAS.449.1819C} {449, 1819}

\bibitem[\protect\citeauthoryear{{Dale}, {Bonnell}  \& {Whitworth}}{{Dale}
  et~al.}{2007}]{DaleJetal2007}
{Dale} J.~E.,  {Bonnell} I.~A.,   {Whitworth} A.~P.,  2007, \mn@doi [\mnras]
  {10.1111/j.1365-2966.2006.11368.x}, \href
  {https://ui.adsabs.harvard.edu/abs/2007MNRAS.375.1291D} {375, 1291}

\bibitem[\protect\citeauthoryear{{Dale}, {W{\"u}nsch}, {Whitworth}  \&
  {Palou{\v{s}}}}{{Dale} et~al.}{2009}]{DaleJetal2009}
{Dale} J.~E.,  {W{\"u}nsch} R.,  {Whitworth} A.,   {Palou{\v{s}}} J.,  2009,
  \mn@doi [\mnras] {10.1111/j.1365-2966.2009.15213.x}, \href
  {https://ui.adsabs.harvard.edu/abs/2009MNRAS.398.1537D} {398, 1537}

\bibitem[\protect\citeauthoryear{{Dale}, {W{\"u}nsch}, {Smith}, {Whitworth}  \&
  {Palou{\v{s}}}}{{Dale} et~al.}{2011}]{DaleJetal2011}
{Dale} J.~E.,  {W{\"u}nsch} R.,  {Smith} R.~J.,  {Whitworth} A.,
  {Palou{\v{s}}} J.,  2011, \mn@doi [\mnras]
  {10.1111/j.1365-2966.2010.17844.x}, \href
  {https://ui.adsabs.harvard.edu/abs/2011MNRAS.411.2230D} {411, 2230}

\bibitem[\protect\citeauthoryear{{Deharveng}, {Lefloch}, {Zavagno}, {Caplan},
  {Whitworth}, {Nadeau}  \& {Mart{\'\i}n}}{{Deharveng}
  et~al.}{2003}]{DeharvengLetal2003}
{Deharveng} L.,  {Lefloch} B.,  {Zavagno} A.,  {Caplan} J.,  {Whitworth} A.~P.,
   {Nadeau} D.,   {Mart{\'\i}n} S.,  2003, \mn@doi [\aap]
  {10.1051/0004-6361:20031157}, \href
  {https://ui.adsabs.harvard.edu/abs/2003A&A...408L..25D} {408, L25}

\bibitem[\protect\citeauthoryear{{Deharveng} et~al.,}{{Deharveng}
  et~al.}{2010}]{DeharvengLetal2010}
{Deharveng} L.,  et~al., 2010, \mn@doi [\aap] {10.1051/0004-6361/201014422},
  \href {https://ui.adsabs.harvard.edu/abs/2010A&A...523A...6D} {523, A6}

\bibitem[\protect\citeauthoryear{{Deharveng} et~al.,}{{Deharveng}
  et~al.}{2015}]{DeharvengLetal2015}
{Deharveng} L.,  et~al., 2015, \mn@doi [\aap] {10.1051/0004-6361/201423835},
  \href {https://ui.adsabs.harvard.edu/abs/2015A&A...582A...1D} {582, A1}

\bibitem[\protect\citeauthoryear{{Dewangan}, {Devaraj}  \& {Ojha}}{{Dewangan}
  et~al.}{2018}]{DewanganLetal2018}
{Dewangan} L.~K.,  {Devaraj} R.,   {Ojha} D.~K.,  2018, \mn@doi [\apj]
  {10.3847/1538-4357/aaaa6f}, \href
  {https://ui.adsabs.harvard.edu/abs/2018ApJ...854..106D} {854, 106}

\bibitem[\protect\citeauthoryear{{Dinnbier}, {W{\"u}nsch}, {Whitworth}  \&
  {Palou{\v{s}}}}{{Dinnbier} et~al.}{2017}]{DinnbierFetal2017}
{Dinnbier} F.,  {W{\"u}nsch} R.,  {Whitworth} A.~P.,   {Palou{\v{s}}} J.,
  2017, \mn@doi [\mnras] {10.1093/mnras/stw3354}, \href
  {https://ui.adsabs.harvard.edu/abs/2017MNRAS.466.4423D} {466, 4423}

\bibitem[\protect\citeauthoryear{{Dobashi}, {Shimoikura}, {Katakura},
  {Nakamura}  \& {Shimajiri}}{{Dobashi} et~al.}{2019}]{DobashiKetal2019}
{Dobashi} K.,  {Shimoikura} T.,  {Katakura} S.,  {Nakamura} F.,   {Shimajiri}
  Y.,  2019, \mn@doi [\pasj] {10.1093/pasj/psz041}, \href
  {https://ui.adsabs.harvard.edu/abs/2019PASJ...71S..12D} {71, S12}

\bibitem[\protect\citeauthoryear{{Dobbs} \& {Wurster}}{{Dobbs} \&
  {Wurster}}{2021}]{DobbsCWursterJ2021}
{Dobbs} C.~L.,  {Wurster} J.,  2021, \mn@doi [\mnras] {10.1093/mnras/stab150},
  \href {https://ui.adsabs.harvard.edu/abs/2021MNRAS.502.2285D} {502, 2285}

\bibitem[\protect\citeauthoryear{{Dobbs}, {Liow}  \& {Rieder}}{{Dobbs}
  et~al.}{2020}]{DobbsCetal2020}
{Dobbs} C.~L.,  {Liow} K.~Y.,   {Rieder} S.,  2020, \mn@doi [\mnras]
  {10.1093/mnrasl/slaa072}, \href
  {https://ui.adsabs.harvard.edu/abs/2020MNRAS.496L...1D} {496, L1}

\bibitem[\protect\citeauthoryear{{Doi} et~al.,}{{Doi}
  et~al.}{2020}]{YasuoDetal2020}
{Doi} Y.,  et~al., 2020, \mn@doi [\apj] {10.3847/1538-4357/aba1e2}, \href
  {https://ui.adsabs.harvard.edu/abs/2020ApJ...899...28D} {899, 28}

\bibitem[\protect\citeauthoryear{{Elmegreen}}{{Elmegreen}}{1994}]{ElmegreenB1994}
{Elmegreen} B.~G.,  1994, \mn@doi [\apj] {10.1086/174147}, \href
  {https://ui.adsabs.harvard.edu/abs/1994ApJ...427..384E} {427, 384}

\bibitem[\protect\citeauthoryear{{Elmegreen} \& {Elmegreen}}{{Elmegreen} \&
  {Elmegreen}}{1978}]{ElmegreeBGDM1978}
{Elmegreen} B.~G.,  {Elmegreen} D.~M.,  1978, \mn@doi [\apj] {10.1086/155991},
  \href {https://ui.adsabs.harvard.edu/abs/1978ApJ...220.1051E} {220, 1051}

\bibitem[\protect\citeauthoryear{{Elmegreen} \& {Lada}}{{Elmegreen} \&
  {Lada}}{1977}]{ElmegreenLada1977}
{Elmegreen} B.~G.,  {Lada} C.~J.,  1977, \mn@doi [\apj] {10.1086/155302}, \href
  {https://ui.adsabs.harvard.edu/abs/1977ApJ...214..725E} {214, 725}

\bibitem[\protect\citeauthoryear{{Enokiya} et~al.,}{{Enokiya}
  et~al.}{2021}]{EnokiyaRetal2021}
{Enokiya} R.,  et~al., 2021, \mn@doi [\pasj] {10.1093/pasj/psaa049}, \href
  {https://ui.adsabs.harvard.edu/abs/2021PASJ...73S.256E} {73, S256}

\bibitem[\protect\citeauthoryear{{Figueira} et~al.,}{{Figueira}
  et~al.}{2017}]{FigueiraMetal2017}
{Figueira} M.,  et~al., 2017, \mn@doi [\aap] {10.1051/0004-6361/201629379},
  \href {https://ui.adsabs.harvard.edu/abs/2017A&A...600A..93F} {600, A93}

\bibitem[\protect\citeauthoryear{{Fissel} et~al.,}{{Fissel}
  et~al.}{2019}]{FisselLetal2019}
{Fissel} L.~M.,  et~al., 2019, \mn@doi [\apj] {10.3847/1538-4357/ab1eb0}, \href
  {https://ui.adsabs.harvard.edu/abs/2019ApJ...878..110F} {878, 110}

\bibitem[\protect\citeauthoryear{{Fukui} et~al.,}{{Fukui}
  et~al.}{2014}]{FukuiYetal2014}
{Fukui} Y.,  et~al., 2014, \mn@doi [\apj] {10.1088/0004-637X/780/1/36}, \href
  {https://ui.adsabs.harvard.edu/abs/2014ApJ...780...36F} {780, 36}

\bibitem[\protect\citeauthoryear{{Girichidis}, {Seifried}, {Naab}, {Peters},
  {Walch}, {W{\"u}nsch}, {Glover}  \& {Klessen}}{{Girichidis}
  et~al.}{2018}]{GirichidisPetal2018}
{Girichidis} P.,  {Seifried} D.,  {Naab} T.,  {Peters} T.,  {Walch} S.,
  {W{\"u}nsch} R.,  {Glover} S. C.~O.,   {Klessen} R.~S.,  2018, \mn@doi
  [\mnras] {10.1093/mnras/sty2016}, \href
  {https://ui.adsabs.harvard.edu/abs/2018MNRAS.480.3511G} {480, 3511}

\bibitem[\protect\citeauthoryear{{G{\'o}mez}, {V{\'a}zquez-Semadeni}  \&
  {Zamora-Avil{\'e}s}}{{G{\'o}mez} et~al.}{2018}]{GomezGetal2018}
{G{\'o}mez} G.~C.,  {V{\'a}zquez-Semadeni} E.,   {Zamora-Avil{\'e}s} M.,  2018,
  \mn@doi [\mnras] {10.1093/mnras/sty2018}, \href
  {https://ui.adsabs.harvard.edu/abs/2018MNRAS.480.2939G} {480, 2939}

\bibitem[\protect\citeauthoryear{{Hayashi} et~al.,}{{Hayashi}
  et~al.}{2021}]{HayashiKetal2021}
{Hayashi} K.,  et~al., 2021, \mn@doi [\pasj] {10.1093/pasj/psaa054}, \href
  {https://ui.adsabs.harvard.edu/abs/2021PASJ...73S.321H} {73, S321}

\bibitem[\protect\citeauthoryear{{Heiles} \& {Troland}}{{Heiles} \&
  {Troland}}{2005}]{HeilesCTrolandT2005}
{Heiles} C.,  {Troland} T.~H.,  2005, \mn@doi [\apj] {10.1086/428896}, \href
  {https://ui.adsabs.harvard.edu/abs/2005ApJ...624..773H} {624, 773}

\bibitem[\protect\citeauthoryear{{Heitsch}, {Slyz}, {Devriendt}, {Hartmann}  \&
  {Burkert}}{{Heitsch} et~al.}{2006}]{HeitschFetal2006}
{Heitsch} F.,  {Slyz} A.~D.,  {Devriendt} J. E.~G.,  {Hartmann} L.~W.,
  {Burkert} A.,  2006, \mn@doi [\apj] {10.1086/505931}, \href
  {https://ui.adsabs.harvard.edu/abs/2006ApJ...648.1052H} {648, 1052}

\bibitem[\protect\citeauthoryear{{Heitsch}, {Hartmann}  \& {Burkert}}{{Heitsch}
  et~al.}{2008}]{HeitschFetal2008}
{Heitsch} F.,  {Hartmann} L.~W.,   {Burkert} A.,  2008, \mn@doi [\apj]
  {10.1086/589919}, \href
  {https://ui.adsabs.harvard.edu/abs/2008ApJ...683..786H} {683, 786}

\bibitem[\protect\citeauthoryear{{Hennebelle} \& {Audit}}{{Hennebelle} \&
  {Audit}}{2007}]{HennebellePAuditE2007}
{Hennebelle} P.,  {Audit} E.,  2007, \mn@doi [\aap]
  {10.1051/0004-6361:20066139}, \href
  {https://ui.adsabs.harvard.edu/abs/2007A&A...465..431H} {465, 431}

\bibitem[\protect\citeauthoryear{{Hennebelle}, {Audit}  \&
  {Miville-Desch{\^e}nes}}{{Hennebelle} et~al.}{2007}]{HennebellePetal2007}
{Hennebelle} P.,  {Audit} E.,   {Miville-Desch{\^e}nes} M.~A.,  2007, \mn@doi
  [\aap] {10.1051/0004-6361:20066141}, \href
  {https://ui.adsabs.harvard.edu/abs/2007A&A...465..445H} {465, 445}

\bibitem[\protect\citeauthoryear{{Inoue} \& {Fukui}}{{Inoue} \&
  {Fukui}}{2013}]{InoueTFukuiY2013}
{Inoue} T.,  {Fukui} Y.,  2013, \mn@doi [\apjl] {10.1088/2041-8205/774/2/L31},
  \href {https://ui.adsabs.harvard.edu/abs/2013ApJ...774L..31I} {774, L31}

\bibitem[\protect\citeauthoryear{{Inoue}, {Hennebelle}, {Fukui}, {Matsumoto},
  {Iwasaki}  \& {Inutsuka}}{{Inoue} et~al.}{2018}]{InoueTetal2018}
{Inoue} T.,  {Hennebelle} P.,  {Fukui} Y.,  {Matsumoto} T.,  {Iwasaki} K.,
  {Inutsuka} S.-i.,  2018, \mn@doi [\pasj] {10.1093/pasj/psx089}, \href
  {https://ui.adsabs.harvard.edu/abs/2018PASJ...70S..53I} {70, S53}

\bibitem[\protect\citeauthoryear{{Kabanovic} et~al.,}{{Kabanovic}
  et~al.}{2022}]{KabanovicSetal2022}
{Kabanovic} S.,  et~al., 2022, \mn@doi [\aap] {10.1051/0004-6361/202142575},
  \href {https://ui.adsabs.harvard.edu/abs/2022A&A...659A..36K} {659, A36}

\bibitem[\protect\citeauthoryear{{Kahn}}{{Kahn}}{1954}]{KahnFD1954}
{Kahn} F.~D.,  1954, \bain, \href
  {https://ui.adsabs.harvard.edu/abs/1954BAN....12..187K} {12, 187}

\bibitem[\protect\citeauthoryear{{Kohno} et~al.,}{{Kohno}
  et~al.}{2018}]{KohnoMetal2018}
{Kohno} M.,  et~al., 2018, \mn@doi [\pasj] {10.1093/pasj/psx137}, \href
  {https://ui.adsabs.harvard.edu/abs/2018PASJ...70S..50K} {70, S50}

\bibitem[\protect\citeauthoryear{{Kwon} et~al.,}{{Kwon}
  et~al.}{2022}]{KwonWetal2022}
{Kwon} W.,  et~al., 2022, \mn@doi [\apj] {10.3847/1538-4357/ac4bbe}, \href
  {https://ui.adsabs.harvard.edu/abs/2022ApJ...926..163K} {926, 163}

\bibitem[\protect\citeauthoryear{{Lee}, {Seon}  \& {Jo}}{{Lee}
  et~al.}{2015}]{DukhangLetal2015}
{Lee} D.,  {Seon} K.-I.,   {Jo} Y.-S.,  2015, \mn@doi [\apj]
  {10.1088/0004-637X/806/2/274}, \href
  {https://ui.adsabs.harvard.edu/abs/2015ApJ...806..274L} {806, 274}

\bibitem[\protect\citeauthoryear{{Li}, {Wang}, {Zhang}, {Ma}  \& {Lin}}{{Li}
  et~al.}{2020}]{LiCetal2020}
{Li} C.,  {Wang} H.,  {Zhang} M.,  {Ma} Y.,   {Lin} L.,  2020, \mn@doi [\apjs]
  {10.3847/1538-4365/ab9cc0}, \href
  {https://ui.adsabs.harvard.edu/abs/2020ApJS..249...27L} {249, 27}

\bibitem[\protect\citeauthoryear{{Liow} \& {Dobbs}}{{Liow} \&
  {Dobbs}}{2020}]{LiowKDobbsC2020}
{Liow} K.~Y.,  {Dobbs} C.~L.,  2020, \mn@doi [\mnras] {10.1093/mnras/staa2857},
  \href {https://ui.adsabs.harvard.edu/abs/2020MNRAS.499.1099L} {499, 1099}

\bibitem[\protect\citeauthoryear{{Mackey}, {Haworth}, {Gvaramadze}, {Mohamed},
  {Langer}  \& {Harries}}{{Mackey} et~al.}{2016}]{MackeyJetal2016}
{Mackey} J.,  {Haworth} T.~J.,  {Gvaramadze} V.~V.,  {Mohamed} S.,  {Langer}
  N.,   {Harries} T.~J.,  2016, \mn@doi [\aap] {10.1051/0004-6361/201527569},
  \href {https://ui.adsabs.harvard.edu/abs/2016A&A...586A.114M} {586, A114}

\bibitem[\protect\citeauthoryear{{Marsh} \& {Whitworth}}{{Marsh} \&
  {Whitworth}}{2019}]{MarshKWhitworthA2019}
{Marsh} K.~A.,  {Whitworth} A.~P.,  2019, \mn@doi [\mnras]
  {10.1093/mnras/sty3186}, \href
  {https://ui.adsabs.harvard.edu/abs/2019MNRAS.483..352M} {483, 352}

\bibitem[\protect\citeauthoryear{{Mestel}}{{Mestel}}{1965}]{MestelL1965b}
{Mestel} L.,  1965, \qjras, \href
  {https://ui.adsabs.harvard.edu/abs/1965QJRAS...6..265M} {6, 265}

\bibitem[\protect\citeauthoryear{{Ohama} et~al.,}{{Ohama}
  et~al.}{2018}]{OhamaAetal2018}
{Ohama} A.,  et~al., 2018, \mn@doi [\pasj] {10.1093/pasj/psy025}, \href
  {https://ui.adsabs.harvard.edu/abs/2018PASJ...70S..45O} {70, S45}

\bibitem[\protect\citeauthoryear{{Palmeirim} et~al.,}{{Palmeirim}
  et~al.}{2017}]{PalmeirimPetal2017}
{Palmeirim} P.,  et~al., 2017, \mn@doi [\aap] {10.1051/0004-6361/201629963},
  \href {https://ui.adsabs.harvard.edu/abs/2017A&A...605A..35P} {605, A35}

\bibitem[\protect\citeauthoryear{{Pattle}, {Fissel}, {Tahani}, {Liu}  \&
  {Ntormousi}}{{Pattle} et~al.}{2022}]{PattleKetal2022}
{Pattle} K.,  {Fissel} L.,  {Tahani} M.,  {Liu} T.,   {Ntormousi} E.,  2022,
  arXiv e-prints, \href {https://ui.adsabs.harvard.edu/abs/2022arXiv220311179P}
  {p. arXiv:2203.11179}

\bibitem[\protect\citeauthoryear{{Peretto} et~al.,}{{Peretto}
  et~al.}{2013}]{PerettoNetal2013}
{Peretto} N.,  et~al., 2013, \mn@doi [\aap] {10.1051/0004-6361/201321318},
  \href {https://ui.adsabs.harvard.edu/abs/2013A&A...555A.112P} {555, A112}

\bibitem[\protect\citeauthoryear{{Peretto} et~al.,}{{Peretto}
  et~al.}{2014}]{PerettoNetal2014}
{Peretto} N.,  et~al., 2014, \mn@doi [\aap] {10.1051/0004-6361/201322172},
  \href {https://ui.adsabs.harvard.edu/abs/2014A&A...561A..83P} {561, A83}

\bibitem[\protect\citeauthoryear{{Pillai} et~al.,}{{Pillai}
  et~al.}{2020}]{PillaiTetal2020}
{Pillai} T. G.~S.,  et~al., 2020, \mn@doi [Nature Astronomy]
  {10.1038/s41550-020-1172-6}, \href
  {https://ui.adsabs.harvard.edu/abs/2020NatAs...4.1195P} {4, 1195}

\bibitem[\protect\citeauthoryear{{Pon}, {Johnstone}  \& {Heitsch}}{{Pon}
  et~al.}{2011}]{PonAetal2011}
{Pon} A.,  {Johnstone} D.,   {Heitsch} F.,  2011, \mn@doi [\apj]
  {10.1088/0004-637X/740/2/88}, \href
  {https://ui.adsabs.harvard.edu/abs/2011ApJ...740...88P} {740, 88}

\bibitem[\protect\citeauthoryear{{Pon}, {Toal{\'a}}, {Johnstone},
  {V{\'a}zquez-Semadeni}, {Heitsch}  \& {G{\'o}mez}}{{Pon}
  et~al.}{2012}]{PonAetal2012}
{Pon} A.,  {Toal{\'a}} J.~A.,  {Johnstone} D.,  {V{\'a}zquez-Semadeni} E.,
  {Heitsch} F.,   {G{\'o}mez} G.~C.,  2012, \mn@doi [\apj]
  {10.1088/0004-637X/756/2/145}, \href
  {https://ui.adsabs.harvard.edu/abs/2012ApJ...756..145P} {756, 145}

\bibitem[\protect\citeauthoryear{{Sakre}, {Habe}, {Pettitt}  \&
  {Okamoto}}{{Sakre} et~al.}{2021}]{SakreNetal2021}
{Sakre} N.,  {Habe} A.,  {Pettitt} A.~R.,   {Okamoto} T.,  2021, \mn@doi
  [\pasj] {10.1093/pasj/psaa059}, \href
  {https://ui.adsabs.harvard.edu/abs/2021PASJ...73S.385S} {73, S385}

\bibitem[\protect\citeauthoryear{{Samal}, {Deharveng}, {Zavagno}, {Anderson},
  {Molinari}  \& {Russeil}}{{Samal} et~al.}{2018}]{SamalMetal2018}
{Samal} M.~R.,  {Deharveng} L.,  {Zavagno} A.,  {Anderson} L.~D.,  {Molinari}
  S.,   {Russeil} D.,  2018, \mn@doi [\aap] {10.1051/0004-6361/201833015},
  \href {https://ui.adsabs.harvard.edu/abs/2018A&A...617A..67S} {617, A67}

\bibitem[\protect\citeauthoryear{{Scoville}, {Sanders}  \&
  {Clemens}}{{Scoville} et~al.}{1986}]{ScovilleNetal1986}
{Scoville} N.~Z.,  {Sanders} D.~B.,   {Clemens} D.~P.,  1986, \mn@doi [\apjl]
  {10.1086/184785}, \href
  {https://ui.adsabs.harvard.edu/abs/1986ApJ...310L..77S} {310, L77}

\bibitem[\protect\citeauthoryear{{Seifried} \& {Walch}}{{Seifried} \&
  {Walch}}{2015}]{SeifriedDWalchS2015}
{Seifried} D.,  {Walch} S.,  2015, \mn@doi [\mnras] {10.1093/mnras/stv1458},
  \href {https://ui.adsabs.harvard.edu/abs/2015MNRAS.452.2410S} {452, 2410}

\bibitem[\protect\citeauthoryear{{Shima}, {Tasker}, {Federrath}  \&
  {Habe}}{{Shima} et~al.}{2018}]{ShimaKetal2018}
{Shima} K.,  {Tasker} E.~J.,  {Federrath} C.,   {Habe} A.,  2018, \mn@doi
  [\pasj] {10.1093/pasj/psx124}, \href
  {https://ui.adsabs.harvard.edu/abs/2018PASJ...70S..54S} {70, S54}

\bibitem[\protect\citeauthoryear{{Soam} et~al.,}{{Soam}
  et~al.}{2019}]{SoamAetal2019}
{Soam} A.,  et~al., 2019, \mn@doi [\apj] {10.3847/1538-4357/ab39dd}, \href
  {https://ui.adsabs.harvard.edu/abs/2019ApJ...883...95S} {883, 95}

\bibitem[\protect\citeauthoryear{{Soler}, {Hennebelle}, {Martin},
  {Miville-Desch{\^e}nes}, {Netterfield}  \& {Fissel}}{{Soler}
  et~al.}{2013}]{SolerJetal2013}
{Soler} J.~D.,  {Hennebelle} P.,  {Martin} P.~G.,  {Miville-Desch{\^e}nes}
  M.~A.,  {Netterfield} C.~B.,   {Fissel} L.~M.,  2013, \mn@doi [\apj]
  {10.1088/0004-637X/774/2/128}, \href
  {https://ui.adsabs.harvard.edu/abs/2013ApJ...774..128S} {774, 128}

\bibitem[\protect\citeauthoryear{{Takahira}, {Tasker}  \& {Habe}}{{Takahira}
  et~al.}{2014}]{TakahiraKetal2014}
{Takahira} K.,  {Tasker} E.~J.,   {Habe} A.,  2014, \mn@doi [\apj]
  {10.1088/0004-637X/792/1/63}, \href
  {https://ui.adsabs.harvard.edu/abs/2014ApJ...792...63T} {792, 63}

\bibitem[\protect\citeauthoryear{{Takahira}, {Shima}, {Habe}  \&
  {Tasker}}{{Takahira} et~al.}{2018}]{TakahiraKetal2018}
{Takahira} K.,  {Shima} K.,  {Habe} A.,   {Tasker} E.~J.,  2018, \mn@doi
  [\pasj] {10.1093/pasj/psy011}, \href
  {https://ui.adsabs.harvard.edu/abs/2018PASJ...70S..58T} {70, S58}

\bibitem[\protect\citeauthoryear{{Tan}}{{Tan}}{2000}]{TanJ2000}
{Tan} J.~C.,  2000, \mn@doi [\apj] {10.1086/308905}, \href
  {https://ui.adsabs.harvard.edu/abs/2000ApJ...536..173T} {536, 173}

\bibitem[\protect\citeauthoryear{{Tanvir} \& {Dale}}{{Tanvir} \&
  {Dale}}{2020}]{TanvirTDaleJ2020}
{Tanvir} T.~S.,  {Dale} J.~E.,  2020, \mn@doi [\mnras] {10.1093/mnras/staa665},
  \href {https://ui.adsabs.harvard.edu/abs/2020MNRAS.494..246T} {494, 246}

\bibitem[\protect\citeauthoryear{{Thompson}, {Urquhart}, {Moore}  \&
  {Morgan}}{{Thompson} et~al.}{2012}]{ThompsonMAetal2012}
{Thompson} M.~A.,  {Urquhart} J.~S.,  {Moore} T.~J.~T.,   {Morgan} L.~K.,
  2012, \mn@doi [\mnras] {10.1111/j.1365-2966.2011.20315.x}, \href
  {https://ui.adsabs.harvard.edu/abs/2012MNRAS.421..408T} {421, 408}

\bibitem[\protect\citeauthoryear{{Walch}, {Whitworth}, {Bisbas}, {Hubber}  \&
  {W{\"u}nsch}}{{Walch} et~al.}{2015}]{WalchSetal2015}
{Walch} S.,  {Whitworth} A.~P.,  {Bisbas} T.~G.,  {Hubber} D.~A.,
  {W{\"u}nsch} R.,  2015, \mn@doi [\mnras] {10.1093/mnras/stv1427}, \href
  {https://ui.adsabs.harvard.edu/abs/2015MNRAS.452.2794W} {452, 2794}

\bibitem[\protect\citeauthoryear{{Ward-Thompson} et~al.,}{{Ward-Thompson}
  et~al.}{2017}]{WardThompsonDetal2017}
{Ward-Thompson} D.,  et~al., 2017, \mn@doi [\apj] {10.3847/1538-4357/aa70a0},
  \href {https://ui.adsabs.harvard.edu/abs/2017ApJ...842...66W} {842, 66}

\bibitem[\protect\citeauthoryear{{Wareing}, {Pittard}  \& {Falle}}{{Wareing}
  et~al.}{2017}]{WareingCetal2017}
{Wareing} C.~J.,  {Pittard} J.~M.,   {Falle} S.~A.~E.~G.,  2017, \mn@doi
  [\mnras] {10.1093/mnras/stw2990}, \href
  {https://ui.adsabs.harvard.edu/abs/2017MNRAS.465.2757W} {465, 2757}

\bibitem[\protect\citeauthoryear{{Whitworth} \& {Jaffa}}{{Whitworth} \&
  {Jaffa}}{2018}]{WhitJaff2018}
{Whitworth} A.~P.,  {Jaffa} S.~E.,  2018, \mn@doi [\aap]
  {10.1051/0004-6361/201731871}, \href
  {https://ui.adsabs.harvard.edu/abs/2018A&A...611A..20W} {611, A20}

\bibitem[\protect\citeauthoryear{{Whitworth} \& {Priestley}}{{Whitworth} \&
  {Priestley}}{2021}]{WhitworthPriestley2021}
{Whitworth} A.~P.,  {Priestley} F.~D.,  2021, \mn@doi [\mnras]
  {10.1093/mnras/stab1125}, \href
  {https://ui.adsabs.harvard.edu/abs/2021MNRAS.504.3156W} {504, 3156}

\bibitem[\protect\citeauthoryear{{Whitworth}, {Bhattal}, {Chapman}, {Disney}
  \& {Turner}}{{Whitworth} et~al.}{1994a}]{Whitetal1994a}
{Whitworth} A.~P.,  {Bhattal} A.~S.,  {Chapman} S.~J.,  {Disney} M.~J.,
  {Turner} J.~A.,  1994a, \mn@doi [\mnras] {10.1093/mnras/268.1.291}, \href
  {https://ui.adsabs.harvard.edu/abs/1994MNRAS.268..291W} {268, 291}

\bibitem[\protect\citeauthoryear{{Whitworth}, {Bhattal}, {Chapman}, {Disney}
  \& {Turner}}{{Whitworth} et~al.}{1994b}]{Whitetal1994b}
{Whitworth} A.~P.,  {Bhattal} A.~S.,  {Chapman} S.~J.,  {Disney} M.~J.,
  {Turner} J.~A.,  1994b, \aap, \href
  {https://ui.adsabs.harvard.edu/abs/1994A&A...290..421W} {290, 421}

\bibitem[\protect\citeauthoryear{{Whitworth}, {Lomax}, {Balfour}, {M{\`e}ge},
  {Zavagno}  \& {Deharveng}}{{Whitworth} et~al.}{2018}]{WhitworthAetal2018}
{Whitworth} A.,  {Lomax} O.,  {Balfour} S.,  {M{\`e}ge} P.,  {Zavagno} A.,
  {Deharveng} L.,  2018, \mn@doi [\pasj] {10.1093/pasj/psx134}, \href
  {https://ui.adsabs.harvard.edu/abs/2018PASJ...70S..55W} {70, S55}

\bibitem[\protect\citeauthoryear{{Wu}, {Tan}, {Nakamura}, {Van Loo}, {Christie}
   \& {Collins}}{{Wu} et~al.}{2017a}]{WuBetal2017a}
{Wu} B.,  {Tan} J.~C.,  {Nakamura} F.,  {Van Loo} S.,  {Christie} D.,
  {Collins} D.,  2017a, \mn@doi [\apj] {10.3847/1538-4357/835/2/137}, \href
  {https://ui.adsabs.harvard.edu/abs/2017ApJ...835..137W} {835, 137}

\bibitem[\protect\citeauthoryear{{Wu}, {Tan}, {Christie}, {Nakamura}, {Van Loo}
   \& {Collins}}{{Wu} et~al.}{2017b}]{WuBetal2017b}
{Wu} B.,  {Tan} J.~C.,  {Christie} D.,  {Nakamura} F.,  {Van Loo} S.,
  {Collins} D.,  2017b, \mn@doi [\apj] {10.3847/1538-4357/aa6ffa}, \href
  {https://ui.adsabs.harvard.edu/abs/2017ApJ...841...88W} {841, 88}

\bibitem[\protect\citeauthoryear{{W{\"u}nsch}, {Dale}, {Palou{\v{s}}}  \&
  {Whitworth}}{{W{\"u}nsch} et~al.}{2010}]{WunschRetal2010}
{W{\"u}nsch} R.,  {Dale} J.~E.,  {Palou{\v{s}}} J.,   {Whitworth} A.~P.,  2010,
  \mn@doi [\mnras] {10.1111/j.1365-2966.2010.17045.x}, \href
  {https://ui.adsabs.harvard.edu/abs/2010MNRAS.407.1963W} {407, 1963}

\bibitem[\protect\citeauthoryear{{Xu} et~al.,}{{Xu}
  et~al.}{2017}]{XuJ-Letal2017}
{Xu} J.-L.,  et~al., 2017, \mn@doi [\apj] {10.3847/1538-4357/aa8ee0}, \href
  {https://ui.adsabs.harvard.edu/abs/2017ApJ...849..140X} {849, 140}

\makeatother
\end{thebibliography}


\bsp	
\label{lastpage}
\end{document}